\documentclass[journal]{IEEEtran}

\usepackage{cite}

\usepackage{graphicx}

\usepackage[cmex10]{amsmath}
\usepackage{bm} 

\usepackage{array}

\usepackage{mdwmath}
\usepackage{mdwtab}

\usepackage{eqparbox}

\usepackage[caption=false,font=footnotesize]{subfig}

\usepackage{color}

\ifCLASSOPTIONcaptionsoff
  \usepackage[nomarkers]{endfloat}
 \let\MYoriglatexcaption\caption
 \renewcommand{\caption}[2][\relax]{\MYoriglatexcaption[#2]{#2}}
\fi

\DeclareMathOperator{\asin}{asin}

\begin{document}

\title{A Leaky-Wave Antenna With Controlled Radiation Using a Bianisotropic Huygens' Metasurface}

\author{Elena~Abdo-S\'anchez, ~\IEEEmembership{Member,~IEEE},
				Michael~Chen,
				Ariel~Epstein, ~\IEEEmembership{Member,~IEEE},
        and George V. Eleftheriades, ~\IEEEmembership{Fellow,~IEEE}\\
				\textit{(DOI: 10.1109/TAP.2018.2878082)}, $\copyright$ \textit{2018 IEEE*}			

\thanks{Personal use of this material is permitted. Permission from IEEE must be obtained for all other uses, in any current or future media, including reprinting/republishing this material for advertising or promotional purposes, creating new collective works, for resale or redistribution to servers or lists, or reuse of any copyrighted component of this work in other works}
\thanks{This project has received funding from the European Union's Horizon 2020 research and innovation programme under the Marie Sklodowska-Curie grant agreement No 706334.}
\thanks{E. Abdo-S\'anchez is with the Departamento de Ingenier\'ia de Comunicaciones, E.T.S.I. Telecomunicaci\'on, Universidad de M\'alaga, Andaluc\'ia Tech, Bulevar Louis Pasteur 35, 29010 M\'alaga, Spain (e-mail: elenaabdo@ic.uma.es).}
\thanks{M. Chen and G.V. Eleftheriades are with the Edward S. Rogers Sr. Department of Electrical and Computer Engineering, University of Toronto, Toronto, ON M5S 2E4 Canada (e-mail: mshuo.chen@mail.utoronto.ca, gelefth@ece.utoronto.ca).}
\thanks{A. Epstein is with the Andrew and Erna Viterbi Faculty of Electrical Engineering, Technion - Israel Institute of Technology, Haifa 32000, Israel (e-mail: epsteina@ee.technion.ac.il).}
} 

\markboth{IEEE TRANSACTIONS ON ANTENNAS AND PROPAGATION}%
{Shell \MakeLowercase{\textit{et al.}}: Bare Demo of IEEEtran.cls for Journals}

\maketitle

\begin{abstract}
In this paper, a novel concept of a leaky-wave antenna is proposed, based on the use of Huygens' metasurfaces. It consists of a parallel-plate waveguide in which the top plate is replaced by a bianisotropic metasurface of the Omega type. It is shown that there is an exact solution to transform the guided mode into a leaky-mode with arbitrary control of the constant leakage factor and the pointing direction. Although the solution turns out to be periodic, only one Floquet mode is excited and radiates, even for electrically long periods. Thanks to the intrinsic spurious Floquet mode suppression, broadside radiation can be achieved without any degradation. Simulations with idealized reactance sheets verify the concept. Moreover, physical structures compatible with PCB fabrication have been proposed and designed, considering aspects such as the effect of losses. Finally, experimental results of two prototypes are presented and discussed.

\end{abstract}


\section{Introduction}
Recently, there has been a significant interest in developing low-profile high-directivity antennas that can be easily mounted on platforms, oriented for applications such as satellite communications or automotive radar. Leaky-wave antennas (LWAs) are a promising solution for this purpose. They consist of a traveling-wave structure that leaks power gradually along its length \cite{Ol93}. Unlike corporate-fed arrays, LWAs have simple feeding and, in turn, reduced complexity. They are characterized by a phase constant (which determines the output angle) and a leakage factor (which controls the radiation rate). Many efforts have been made lately to achieve an independent control of these two parameters, such that versatile radiation patterns can be designed at will \cite{MarRos12}. For example, by controlling the leakage factor, high aperture illumination efficiencies can be achieved and, thus, narrower beams for a given antenna length or area.

Traditionally, periodic LWAs have suffered from the problem of the \textit{open-stopband} effect, which deteriorates the radiation characteristics when scanning through broadside. For broadside radiation, the phase shift in a period is a multiple of $2\pi$, which causes all the reflections to add in phase at the source. As a consequence, strong frequency variations of the input impedance are observed around this singular frequency and some mismatch is obtained, which degrades the radiation performance (the amount of radiation drops substantially) \cite{Ol93}. In terms of Floquet mode analysis, this phenomenon at broadside is interpreted as coupling between Floquet modes \cite{He69}. In the past years, LWAs which overcome this problem have been proposed, starting with the metamaterial-inspired Composite Right-Left Handed (CRLH) LWA, which, unlike other periodic LWAs, radiates from the fundamental harmonic \cite{Li02}. Recently, techniques based on circuit theory have been proposed to mitigate the open-stopband phenomenon at broadside in periodic LWAs radiating from the $m=-1$ spatial harmonic as well \cite{Pa09,Ot11,Ot14,Ab16}. 

The appearance of metasurfaces has allowed advanced manipulation of the electromagnetic field, hence providing significant control over the radiation characteristics of planar surfaces \cite{Ho12}. Some examples are LWAs based on modulated metasurfaces (holographic antennas and modulated metasurface antennas) in which a surface wave is transformed to a leaky mode by properly modulating an impedance surface \cite{Fo10,Pa11,Mi11,Mi15}. Although the modulation of the surface allows meticulous control of the leakage factor, achieving high aperture illumination could be challenging \cite{Fa16}. Moreover, some of these LWAs that are fed from an edge experience the open-stopband problem and cannot radiate at broadside \cite{Pa11}. In \cite{Ti15}, a pattern synthesis procedure for LWAs was proposed for a structure consisting of a longitudinally-varying impedance sheet on a grounded dielectric whose permittivity also changes along the longitudinal dimension to get the desired propagation constant. In a subsequent work \cite{Ti18}, the authors show that control of the amplitude, phase and polarization of leaky-wave modes can be achieved using full-tensor stacked electric sheet impedances. However, broadside radiation is not discussed and designs with physical structures are not provided.

Recently, Huygens' metasurfaces have been proposed as a powerful tool for arbitrary wavefront manipulation \cite{Pf13,Mo13,Se13}. They are based on the equivalence principle, which leads to the statement that, given an incident field, an arbitrary aperture field can be achieved by inducing the required electric and (equivalent) magnetic surface currents \cite{Ba05}. These metasurfaces consist of sub-wavelength electrically- and magnetically-polarizable particles that allow the fulfillment of the required boundary conditions to achieve the desired field transformation. In \cite{Ep14}, the authors demonstrated that directive radiation to a prescribed angle when excited by a given (arbitrary) source field can be achieved with lossless and passive particles by satisfying two physical conditions: local power conservation across the surface, and local impedance equalization of the fields on both sides of the metasurface. Although this formulation has allowed the design of enhanced antennas \cite{Ep16_Nat}, its applicability is limited since it does not allow the control of the reflection coefficient, which is mandatory for the design of a LWA with arbitrary choice of the leakage factor.

Nevertheless, it has been recently discovered that by introducing bianisotropy of the Omega type into the particles used for the metasurface implementation, the local impedance equalization condition is not required anymore to obtain passive and lossless metasurfaces, and local power conservation along the metasurface alone suffices to achieve the desired field transformation \cite{Ep16}. This is possible due to the additional (magnetoelectric) degree of freedom provided by the Bianisotropic Huygens' Metasurfaces (BHMSs) of the Omega type. The term `Omega' refers here exclusively to the type of bianisotropy (not to the particle shape), which is given to reciprocal bianisotropic particles in which, unlike chiral particles, the induced electric and magnetic currents have the same polarization as the applied electric and magnetic fields, respectively \cite{Ra13}. The fact that only one condition for arbitrary field transformation is required entails more degrees of freedom for setting the desired output field.

In this contribution, we propose the use of a BHMS as a top plate of a parallel-plate waveguide to build a LWA with arbitrary control of the radiation parameters and extend the preliminary work of \cite{Ab17}. We derive the complete theoretical formulation to obtain the required metasurface parameters to convert the guided mode into a leaky-mode with certain leakage factor and pointing direction. It is shown, for the first time to the authors' knowledge, that there is an exact solution for the boundary problem to convert the guided mode into a single spatial harmonic by means of a periodic lossless and passive metasurface, with no restriction in the period.  
Furthermore, we describe the design methodology to convert the theoretical metasurface parameters into a physical structure compatible with PCB fabrication and discuss implementation aspects. Different designs are shown with physical realizations to validate the theory. Finally, the design cycle is closed with the fabrication of two of the designs and the experimental verification of the concept.

\section{Concept and Theory}\label{s:Theory}
\begin{figure}[!t]
\centering
\includegraphics[width=\columnwidth]{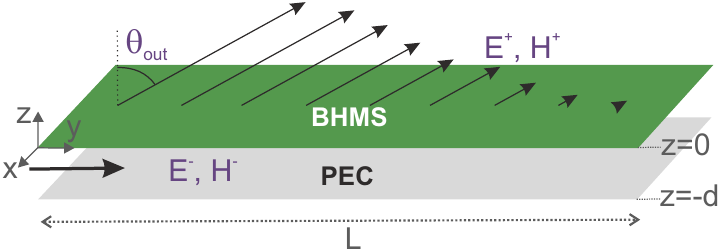}
\caption{Proposed LWA configuration, which consists of a parallel-plate waveguide with the top plate being a bianisotropic metasurface.}
\label{fig:Esquema}
\end{figure}

The proposed structure (Fig. \ref{fig:Esquema}) is a parallel-plate waveguide in which the top plate is replaced by a BHMS of the Omega type. In this way, the BHMS will be the part of the guiding structure leaking power outside. We consider a 2D-configuration ($\partial/\partial x=0$) with the BHMS located at $z=0$ and a perfect electric conductor (PEC) plate located at $z=-d$. The BHMS has a length in the y-coordinate of $L$, and the excitation of the resulting parallel-plate waveguide is located at $y=0$. A transverse electric (TE) polarized field is used as field excitation ($E_y=E_z=H_x=0$). Analogously as done for anisotropic metasurfaces \cite{Ku03}, the bianisotropic sheet transition conditions for scalar Omega-type bianisotropic metasurfaces \cite{Ep16, Ra13} relate the transverse field components above ($E_x^+$ and $H_y^+$) and below ($E_x^-$ and $H_y^-$) the BHMS ($z\rightarrow\pm0$) following
\begin{equation}
\begin{split}
\tfrac{1}{2}(E_x^++E_x^-)=-Z_{se}(H_y^+-H_y^-)-K_{em}(E_x^+-E_x^-)\\
\tfrac{1}{2}(H_y^++H_y^-)=-Y_{sm}(E_x^+-E_x^-)+K_{em}(H_y^+-H_y^-)
\end{split} \label{eq:BSTC}
\end{equation}
where $Z_{se}$ stands for the electric surface impedance, $Y_{sm}$ for the magnetic surface admittance and $K_{em}$ for the magnetoelectric coupling coefficient.

According to \cite{Ep16}, only one condition must be fulfilled if we want to achieve an arbitrary field transformation with passive and lossless particles, i.e. with $\text{Re}[Z_{se}]=\text{Re}[Y_{sm}]=\text{Im}[K_{em}]=0$. This restriction is termed the \textit{local power conservation condition} and implies the conservation of the real power along the perpendicular axis at each point of the metasurface, namely $P_z^-(y)=P_z^+(y)$, where $P_z=\frac{1}{2}\text{Re}\left\{E_xH_y^*\right\}$. From (\ref{eq:BSTC}), the metasurface parameters $Z_{se}$, $Y_{sm}$ and $K_{em}$ can be obtained as a function of $\left\{E_x^+, E_x^-, H_y^+, H_y^-\right\}$ evaluated at $z\rightarrow\pm0$ as follows \cite{Ep16}:
\begin{equation}
\begin{split}
K_{em}&=\tfrac{1}{2}\tfrac{\text{Re}[E_x^+H_y^{-*}-E_x^-H_y^{+*}]}{\text{Re}[(E_x^+-E_x^-)(H_y^+-H_y^-)^*]}\\
Y_{sm}&=-j\left(\tfrac{1}{2}\text{Im}\left[\tfrac{H_y^++H_y^-}{E_x^+-E_x^-}\right]-K_{em}\text{Im}\left[\tfrac{H_y^+-H_y^-}{E_x^+-E_x^-}\right]\right)\\
Z_{se}&=-j\left(\tfrac{1}{2}\text{Im}\left[\tfrac{E_x^++E_x^-}{H_y^+-H_y^-}\right]+K_{em}\text{Im}\left[\tfrac{E_x^+-E_x^-}{H_y^+-H_y^-}\right]\right).
\end{split} \label{eq:MetaParam}
\end{equation}

Therefore, the first step of the theoretical derivation of the problem is to stipulate the fields below and above the metasurface, such that the desirable transformation is achieved, the fields satisfy Maxwell's equations and the relevant boundary conditions, and the local power conservation condition is satisfied along the metasurface. Since the metasurface design will force the boundary conditions at $z=0$, the only restriction for the field below the BHMS is to vanish at the PEC ($z=-d$) and meet the wave equation. Therefore, the following electromagnetic field below the metasurface has been stipulated \cite{Ab17}, corresponding to a leaky guided-wave mode:
\begin{subequations}\label{eq:FieldBelow}
\begin{equation}
\begin{split}
E_x^-=&|E_{in}|(e^{jk_z^-(z+d)}-e^{-jk_z^-(z+d)})e^{-jk_y^-y}\\=&2j|E_{in}|\sin(k_z^-(z+d))e^{-jk_y^-y}
\end{split}
\end{equation}
\begin{equation}
\begin{split}
H_y^-&=\frac{j}{k^-\eta^-}\frac{\partial E_x^-}{\partial z}\\&=-|E_{in}|\frac{k_z^-}{k^-\eta^-}(e^{jk_z^-(z+d)}+e^{-jk_z^-(z+d)})e^{-jk_y^-y}\\
&=-2|E_{in}|\frac{k_z^-}{k^-\eta^-}\cos(k_z^-(z+d))e^{-jk_y^-y}\,.
\end{split}
\end{equation}
\end{subequations}

The desired field for the region above the metasurface is stipulated as a leaky mode, simply as \cite{Ol93} 
\begin{subequations}\label{eq:FieldAbove}
\begin{equation}
E_x^+=|E_{out}|e^{-jk_z^+z}e^{-jk_y^+y}e^{j\xi}
\end{equation}
\begin{equation}
H_y^+=\frac{j}{k^+\eta^+}\frac{\partial E_x^+}{\partial z}=|E_{out}|\frac{k_z^+}{k^+\eta^+}e^{-jk_z^+z}e^{-jk_y^+y}e^{j\xi}\,,
\end{equation}
\end{subequations}
where $\xi$ is a constant phase shift that can be added to the wave as a degree of freedom.

In order to allow the structure to radiate, the propagation constants are complex:
\begin{equation}
\begin{split}
	k_y^-&=\beta^- -j\alpha^-;\ k_z^-=\beta_z^- -j\alpha_z^-; \ {k^-}^2={k_y^-}^2+{k_z^-}^2\\
	k_y^+&=\beta^+ -j\alpha^+;\ 	k_z^+=\beta_z^+ -j\alpha_z^+; \ {k^+}^2={k_y^+}^2+{k_z^+}^2\,,
\end{split}
\end{equation}
where we assume constant $\beta$ and $\alpha$ along $y$. 

The power profiles just below and above the metasurface can be calculated from (\ref{eq:FieldBelow}) and (\ref{eq:FieldAbove}), yielding
\begin{equation}
\begin{split}
	P_z^-(y)&=-\frac{|E_{in}|^2}{\eta^- k^-}\left(\beta_z^-\sinh(2\alpha_z^- d)-\alpha_z^-\sin(2\beta_z^- d)\right)e^{-2\alpha^-y}\\
	P_z^+(y)&=\frac{1}{2}|E_{out}|^2\frac{\beta_z^+}{\eta^+ k^+} e^{-2\alpha^+y}\,.
\end{split}
\end{equation}
Then, for the local power conservation condition to be met, the fields must have the same decay rate along $y$, i.e. $\alpha^+=\alpha^-=\alpha$, where $\alpha$ is defined as the leakage factor. Moreover, the amplitudes of the fields must fulfill the following condition:
\begin{equation}
|E_{out}|=|E_{in}|\sqrt{2\frac{\eta^+k^+}{\eta^-k^-\beta_z^+}(\alpha_z^-\sin(2\beta_z^-d)-\beta_z^-\sinh(2\alpha_z^-d))}\,,\label{s:ConditionEout}
\end{equation}
which does not impose any additional restriction to the fields. 

In fact, 
(\ref{eq:FieldBelow})-(\ref{s:ConditionEout}) demonstrate that there is an exact solution to convert a guided more into a leaky wave in a given direction by means of a lossless and passive metasurface. If we substitute the fields into (\ref{eq:MetaParam}), we can obtain the metasurface parameters $\left\{K_{em}, Z_{se}, Y_{sm}\right\}$ as a function of the propagation constants of the guided and radiated modes and the waveguide height $d$. It can be demonstrated that when $\alpha$ is constant, the resulting metasurface constituents $\left\{K_{em}, Z_{se}, Y_{sm}\right\}$ are periodic, with a period given by
\begin{equation}
p=\frac{2\pi}{\left|\beta^+-\beta^-\right|}\,.\label{eq:period}
\end{equation}

It is well known, according to Floquet's theorem \cite{He69}, that the field scattered off a periodic structure can be expressed in terms of an infinite number of so-called \textit{spatial harmonics}, whose phase constants are given by
\begin{equation}
\beta_m=\beta_0+\frac{2\pi}{p}m\,,\label{eq:FloquetTheorem}
\end{equation}
where $m$ is an integer number indicating the spatial harmonic, and $\beta_0$ is the fundamental phase constant ($\beta^-$ in our case).

One should notice the relation between expressions (\ref{eq:period}) and (\ref{eq:FloquetTheorem}). In fact, it reveals that the radiation from the metasurface will occur through the first higher-order spatial harmonics, +1 or -1 depending on the chosen solution for the denominator in (\ref{eq:period}) ($\beta_{\pm1}=\beta^+$ and $\beta_0=\beta^-$). Therefore, it is worth pointing out that the period obtained from the `blind' theoretical derivation for the aimed field transformation coincides with the period provided by the Floquet's theorem. Nevertheless, according to Floquet's theorem, all the spatial harmonics that fulfill the condition $|\beta_m|<k_0$ would be able to radiate; however, as in our previous works on metasurfaces \cite{Ep14, Ep16, Ep16_2}, our derivation guarantees that only one of these modes carries power. This is an interesting feature, since normally LWA designers choose short periods to make sure that there is only one spatial harmonic that fulfills the radiation condition \cite{Ya10}. In contrast, the methodology presented herein allows operation with long periods since the spurious spatial harmonics will not be excited (this can be interesting in terms of scan rates when making the antenna reconfigurable). Hence, it is hereby shown that there is an exact solution for the boundary problem to convert a guided mode into a single spatial harmonic by means of a periodic lossless and passive metasurface with no restriction in the period. A consequence of the fact of having a single spatial harmonic in the field above the metasurface is the intrinsic suppression of the open-stopband effect when radiating at broadside, since there cannot be any coupling between harmonics \cite{He69}. 

\section{Design and Implementation Methodology}\label{s:DesignImplementation}
The phase constants in the two regions of the space, described in Section \ref{s:Theory}, $\beta^+$ and $\beta^-$, can be related to the pointing angle $\theta_{out}$ and the angle of incidence inside the LWA $\theta_{in}$, respectively, by
\begin{equation}
\begin{split}
\beta^+&\approx k^+ \sin(\theta_{out})\\
\beta^-&\approx k^- \sin(\theta_{in})\,.
\end{split}
\end{equation}

In order to have control of the radiation pattern of the LWA, the pointing angle, $\theta_{out}$, and the leakage factor, $\alpha$, must be chosen independently. As we have seen in the theoretical derivation, both parameters can indeed be set arbitrarily via the presented methodology. Moreover, the metasurface allows decoupling of the waveguiding and radiation problems (similar to \cite{Ep16_Nat}), such that either the period of the structure, $p$, or $\theta_{in}$ are degrees of freedom, together with the waveguide height $d$. Therefore, 
the desired field transformation can be achieved with practically all degrees of freedom possible. In fact, $d$ can be chosen so that the equivalent parallel-plate waveguide (i.e., with both top and bottom metallic plates) would be under cut-off \cite{Ab17}. However, the leaky waveguide formed by the properly designed BHMS will guarantee the propagation of the field as stipulated, as well as the desirable transformation.

Once we have set all the parameters of the desired fields, we must calculate the metasurface constituents $\left\{K_{em}, Z_{se}, Y_{sm}\right\}$, that, according to the derivation, will be purely real ($K_{em}$) and purely imaginary ($Z_{se}$ and $Y_{sm}$), corresponding to passive and lossless particles \cite{Ep16}. Then, in order to obtain a physical implementation of the metasurface, sampling of these parameters is needed, due to the impossibility of building a continuously varying metasurface. This sampling rate will determine the length of the unit-cells (or meta-atoms) and the required number of unit-cells per period. We have chosen a sampling rate of $\frac{\lambda_0}{6}$, a value that has been previously shown to allow the homogenization approximation (e.g., see \cite{Pf14}).

At each point $y=y_0$, a suitable equivalent circuit can be obtained for the corresponding meta-atom by assuming local periodicity (i.e., an infinite array of identical unit-cells). Then, the local properties of the BHMS at each point $y=y_0$ can be approximated by the scattering properties of this infinite periodic array \cite{Ep16_Nat}. In this way, each unit-cell can be characterized by a 2x2 impedance matrix $[\textbf{Z}]$ that relates the tangential fields below and above the metasurface \cite{Se13_b}:
\begin{equation}
\left( \begin{array}{c} E_x^- \\ E_x^+ \end{array} \right) =
\left( \begin{array}{cc} Z_{11}&Z_{12} \\ Z_{21}&Z_{22} \end{array}\right)
\left( \begin{array}{c} H_y^- \\ -H_y^+ \end{array} \right).
\end{equation}

By using (\ref{eq:BSTC}), we can express the impedance matrix in terms of the metasurface parameters \cite{Ep16}:
\begin{equation}
\begin{split}
Z_{11}&=Z_{se}+\frac{(1+2K_{em})^2}{4Y_{sm}}\\
Z_{12}&=Z_{21}=Z_{se}-\frac{(1-2K_{em})(1+2K_{em})}{4Y_{sm}}\\
Z_{22}&=Z_{se}+\frac{(1-2K_{em})^2}{4Y_{sm}}.\\
\end{split}
\end{equation}

\begin{figure}[!t]
\centering{
	\begin{tabular}{cc}
	\subfloat[]{\includegraphics[width=2.5cm]{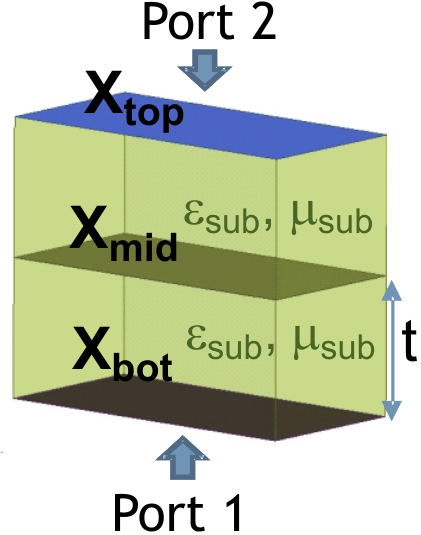}%
	}
	\qquad
	\subfloat[]{\includegraphics[width=5.8cm]{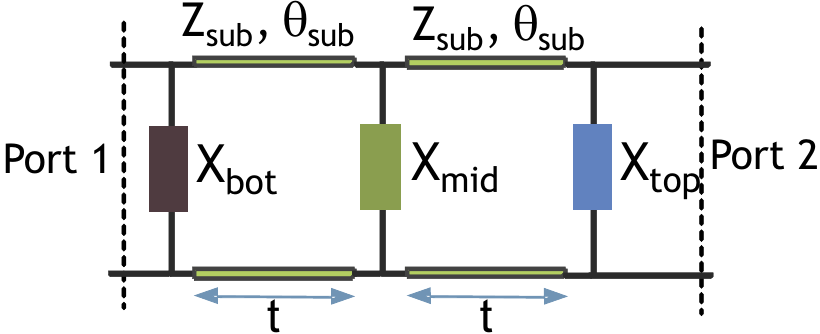}\label{fig:ModelThreeCascadedSheets}
	}
	\end{tabular}}
\caption{(a) Implementation of a BHMS unit-cell with three cascaded reactance sheets. (b) Equivalent circuit of (a).}\label{fig:CascadedSheets}
\end{figure}

It can be noted that, due to the magneto-electric coupling, the Z-Matrix is not symmetric, unlike for non-bianisotropic Huygens' metasurfaces ($K_{em}=0$). However, in the same way, the equivalent circuit can be implemented by three cascaded impedance sheets and closed-form expressions can be straightforwardly derived \cite{Ep16}. The implementation of a general unit-cell with three cascaded impedance sheets can be observed in Fig. \ref{fig:CascadedSheets}. The unit-cell can be seen as a two-port circuit consisting of three parallel reactances connected through transmission lines that represent the dielectrics in the vertical direction \cite{Pf14, Ch18}. 

\section{Physical Realization}\label{s:PhysicalRealization}
There are several challenges regarding the physical realization of the metasurface. First, in the theoretical derivation, lossless particles are used to implement the desired field transformation. However, in practical realizations, losses are inevitable. Typically, the three-layered abstract structure presented in Fig. \ref{fig:CascadedSheets} is implemented via three copper traces defined on a dielectric substrate \cite{Ch18}. The metal forming the trace to realize the desired impedance has losses and so has the substrate over which the trace is printed. Therefore, losses are an undesired effect that will make the theoretical design diverge from the real implementation. Furthermore, unlike other structures (e.g., metasurface for anomalous refraction \cite{Ch18,La18}), the LWA application of bianisotropic Huygens' metasurfaces can require unit-cells with a very high reflection coefficient. This means that the transmitted wave is not impinging on the metasurface just once, but there will be multiple reflections inside the waveguide to guide the mode thus increasing the losses. 

\begin{figure}[!t]
\centering
\includegraphics[width=8cm]{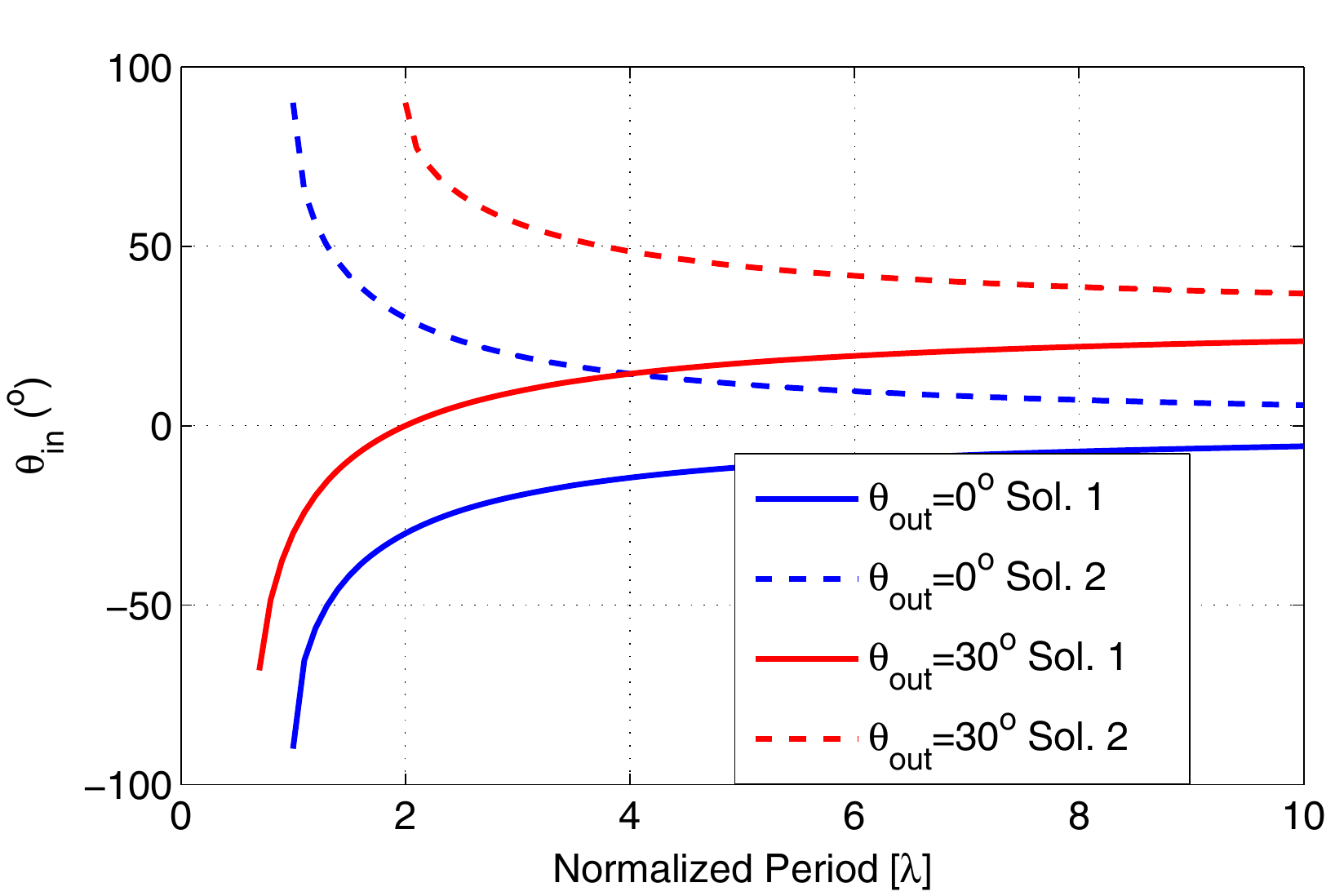}
\caption{Dependence between the period of the metasurface and the input and output angles, $\theta_{in}$ and $\theta_{out}$. The two solutions correspond to the two possible signs of $\left(\beta^+-\beta^-\right)$ in the denominator of (\ref{eq:period}) (Sol. 1 corresponds to the harmonic $m=-1$ and Sol.2 to $m=+1$).}
\label{fig:AnglesAndPeriodDependence}
\end{figure}

It can be understood that a small propagation angle $\theta_{in}$ would lead to much higher losses than a larger $\theta_{in}$ for the same field transformation and type of unit-cells. This is due to the higher number of reflections when the field inside the waveguide is impinging close to normal. Therefore, in order to reduce losses, a large propagation angle of the waveguided mode $\theta_{in}$ (close to the horizon) is desirable. However, as aforementioned, there is a dependence between $\theta_{in}$, $\theta_{out}$ and the period, $p$. Fig. \ref{fig:AnglesAndPeriodDependence} shows this dependence for $\theta_{out}$=0$^\circ$ and $\theta_{out}$=30$^\circ$. The two solutions correspond to the two possible signs of $\beta^+-\beta^-$ in the denominator of (\ref{eq:period}) (Sol. 1 corresponds to the harmonic $m=-1$ and Sol. 2 to $m=+1$). For instance, for $\theta_{out}$=0$^\circ$ it is necessary to have a very short period to get large $\theta_{in}$. In view of practical discretization constraints, this would correspond to a small number of unit cells per period, which might not be enough to model the required variation of the metasurface parameters within a period. Therefore, a trade-off is imposed between the two requirements (avoiding losses while allowing sufficient period size for discretized realization).

Additionally, it is important to bear in mind that, in this work, and unlike other applications of Huygens' metasurfaces, both the phase and the magnitude of the $S_{11}$, $S_{12}$ and $S_{22}$ of each unit-cell (obtained as the two-port represented in Fig. \ref{fig:CascadedSheets} when local periodicity is imposed) must be adjusted for a proper design. Although it was previously shown that any such passive and lossless scattering parameters can be implemented using three cascaded impedance sheets \cite{Mo13,Ep16,Pf14}, these formulations assumed lossless structures; when realistic losses are considered, three degrees of freedom are no longer sufficient for achieving this goal. 
Therefore, similar to \cite{Pf14}, we will employ four layers, so that we have more degrees of freedom to adjust the total scattering parameters of the unit-cells. Having four layers, however, complicates the design and the possible optimization.

\begin{figure}[!t]
\centering
\includegraphics[width=4.5cm]{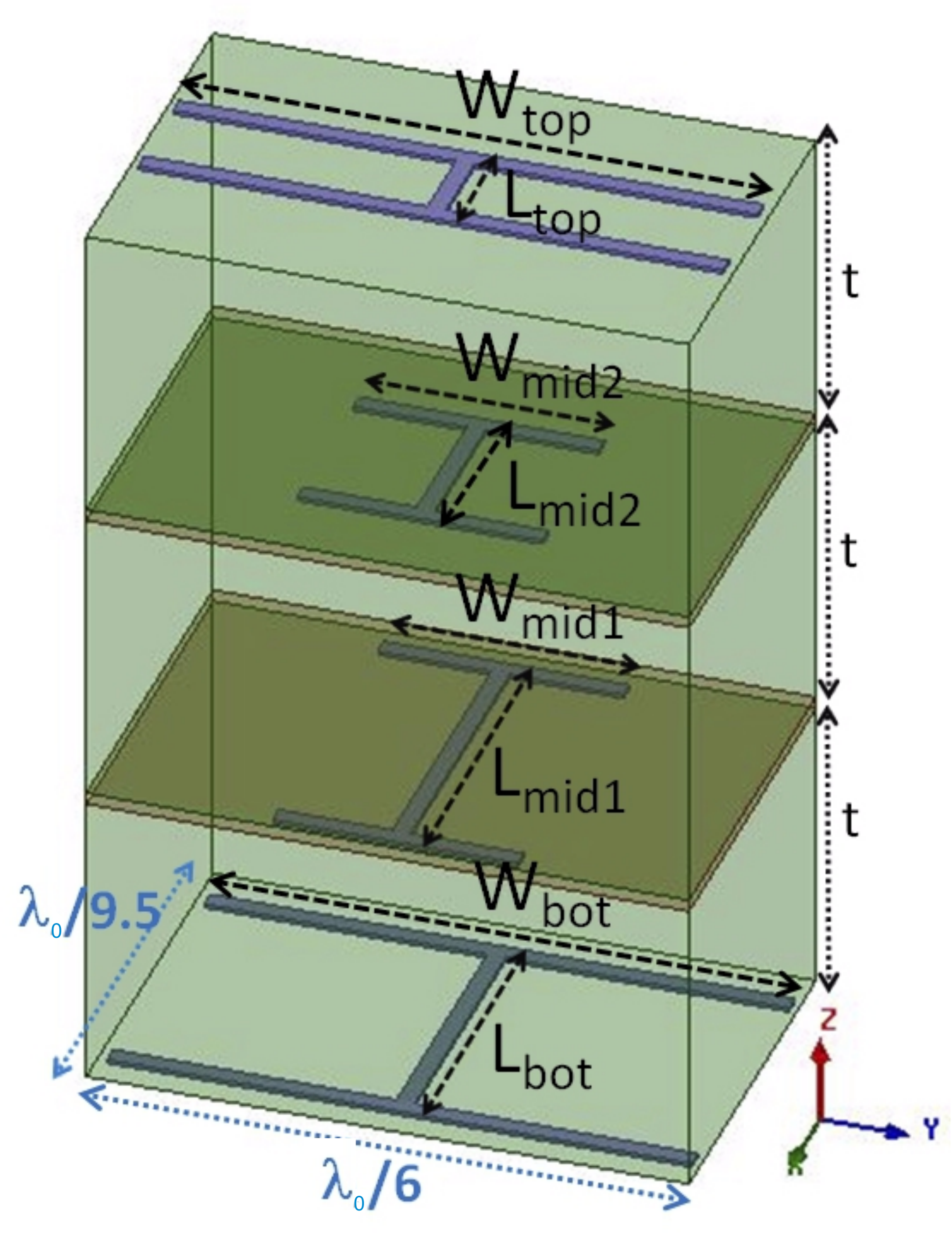}
\caption{Implementation of the unit-cell using dog-bones and four layers. Three dielectric laminates of RO3010 with $\epsilon_r$=12.94 and thickness $t=50$\,mil are bonded with the bondply RO2929, with $\epsilon_r=2.94$ and thickness $t_{BP}=2$\,mil. The copper traces are 18\,$\mu m$ thick and 3\,mil wide.}
\label{fig:RealUnitCell}
\end{figure}

The three-layer solution shown in Fig. \ref{fig:CascadedSheets} utilizes a transmission line (TL) model, assuming negligible near-field interlayer coupling. Thus, to be able to use a similar analytical model with four layers, we should keep the substrate thick enough to avoid significant near-field coupling between the layers. 
The employed $\lambda_0/6\times\lambda_0/9.8\times\lambda_0/3.8$ unit-cell geometry is shown in Fig. \ref{fig:RealUnitCell}. It consists of three laminates of Rogers RO3010 ($\epsilon_r=12.94$ at 20GHz) with thickness $t=50$\,mil bonded with the 2\,mil-thick bondply Rogers RO2929. The bondply is used as a thin film adhesive in high performance, high reliability multi-layer structures, since it has low losses. The electromagnetic properties (including anisotropy) of the RO3010 laminates at 20\,GHz (the chosen frequency of operation, with $\lambda_0\approx15\,mm$) have been used in the electromagnetic simulator ANSYS Electromagnetic Suite 16 (HFSS 2015). The copper traces are 18\,$\mu m$ thick and 3\,mil wide, and the widths of the dog-bones ($W_n$) of each layer have been kept constant for all the unit-cells, thus tuning the lengths ($L_n$) to achieve the desired impedance values.

\begin{figure}[!t]
\centering
\includegraphics[width=6cm]{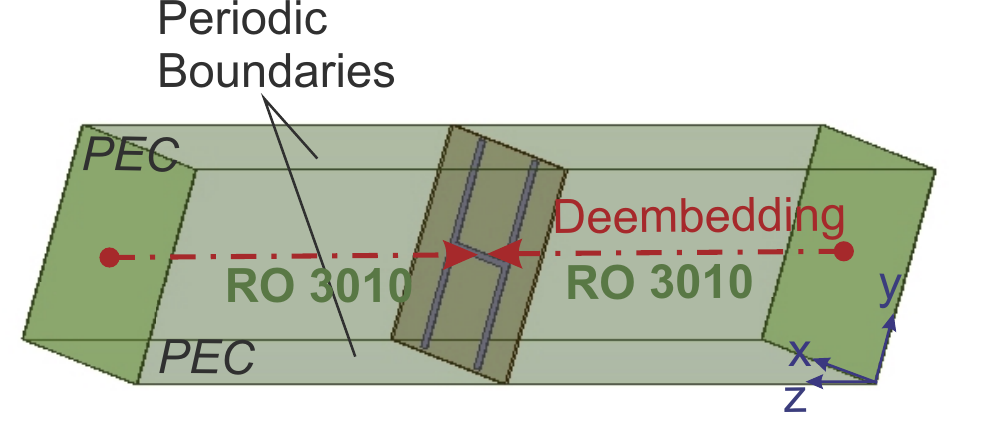}
\caption{Simulation to extract the behavior (ABCD matrix) of the middle layers. The walls in the zy-planes are simulated as PEC, while periodic boundary conditions (master and slave) are applied at the walls in the zx-planes.}
\label{fig:SimCellMidLayer}
\end{figure}

\begin{figure}[!t]
\centering
\includegraphics[width=8cm]{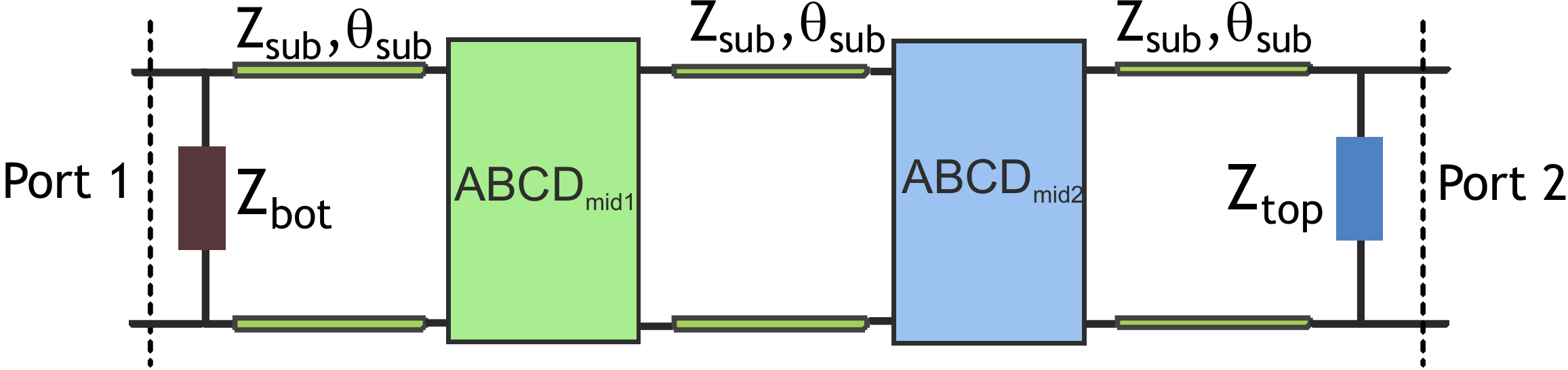}
\caption{Equivalent model of the unit-cell used to calculate the cascaded response.}
\label{fig:ModelRealUnitCell}
\end{figure}

Each layer has been simulated separately (in a similar way as explained in \cite{Pf14}) and the sheet impedance (complex value) as a function of the length of the dog-bone has been extracted. In the case of the middle layers, the presence of the bondply must be taken into account to extract the layer behavior. Then, a simulation of the metallic dogbone surrounded by RO3010 (and extracting the sheet impedance from it by de-embedding) is not appropriate. In order to consider the effect of the dogbone together with the bondply, both surrounded by RO3010 dielectric, we have worked with the ABCD matrix after de-embedding up to the bondply (as illustrated in Fig. \ref{fig:SimCellMidLayer}). The walls in the zy-planes and in the zx-planes are simulated as PEC and periodic boundary conditions, respectively, to simulate an infinite periodic structure. For the bottom and top layer, the widths of the dog-bones were set to $W_{bot}$=$W_{top}$=95.3\,mil ($\approx\lambda_0/6-3$\,mil) and, for the middle layers, $W_{mid1}=W_{mid2}=40$\,mil. These values were chosen to allow a wider range of impedance sheets. Then, the cascaded response for all possible combinations is analytically calculated by using the corresponding TL model, shown in Fig. \ref{fig:ModelRealUnitCell}.

\begin{figure}[!t]
\centering
\includegraphics[width=8cm]{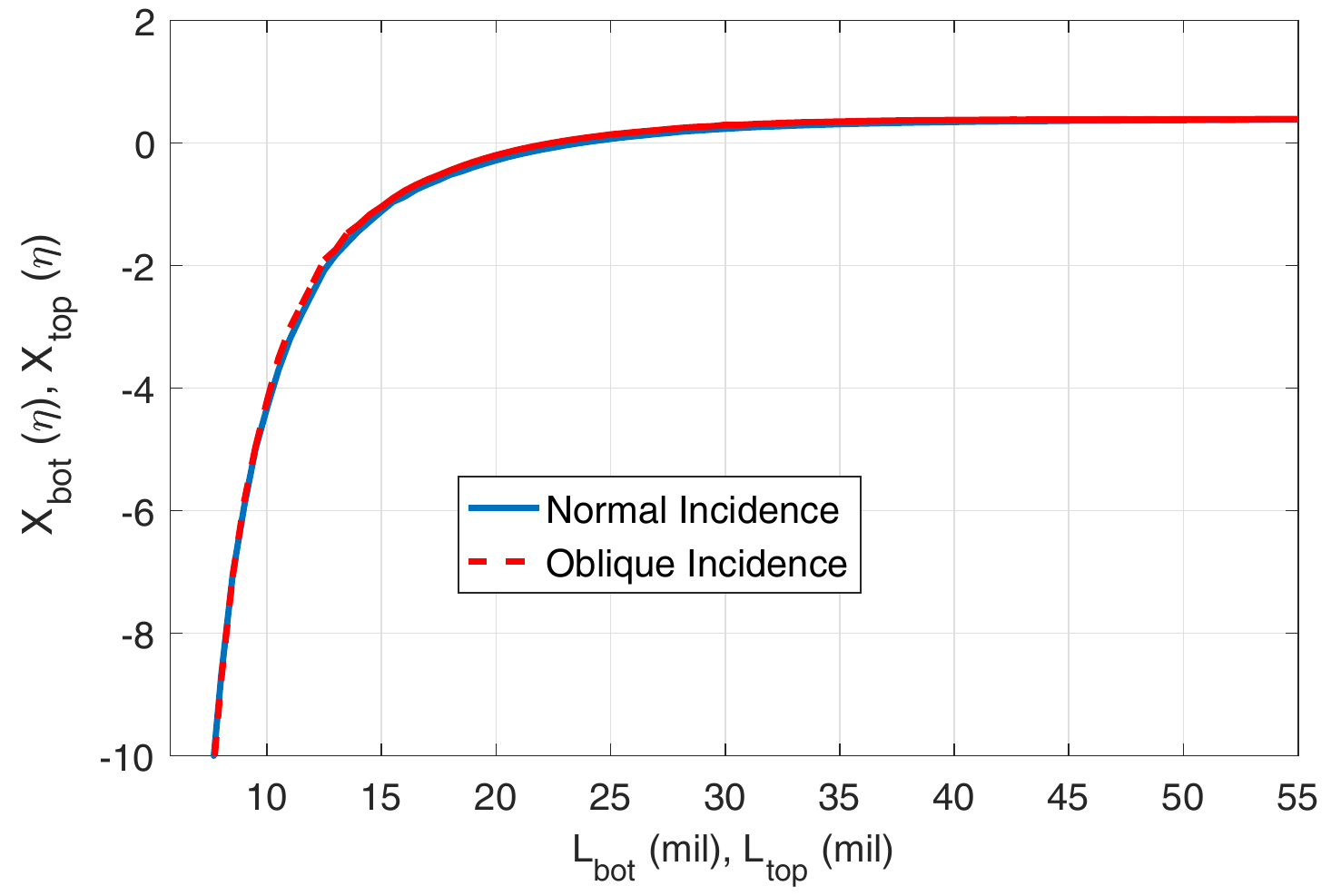}
\caption{Resulting imaginary part of the sheet impedance of the bottom (or top) layer versus the length of the dogbone ($L_{bot}$ or $L_{top}$), when considering normal incidence and oblique incidence with $\theta_{in}=48.6^\circ$.}
\label{fig:Xbottop}
\end{figure}

\begin{figure*}[t]
\centering{
	\begin{tabular}{ccc}
	\subfloat[]{\includegraphics[width=5.7cm]{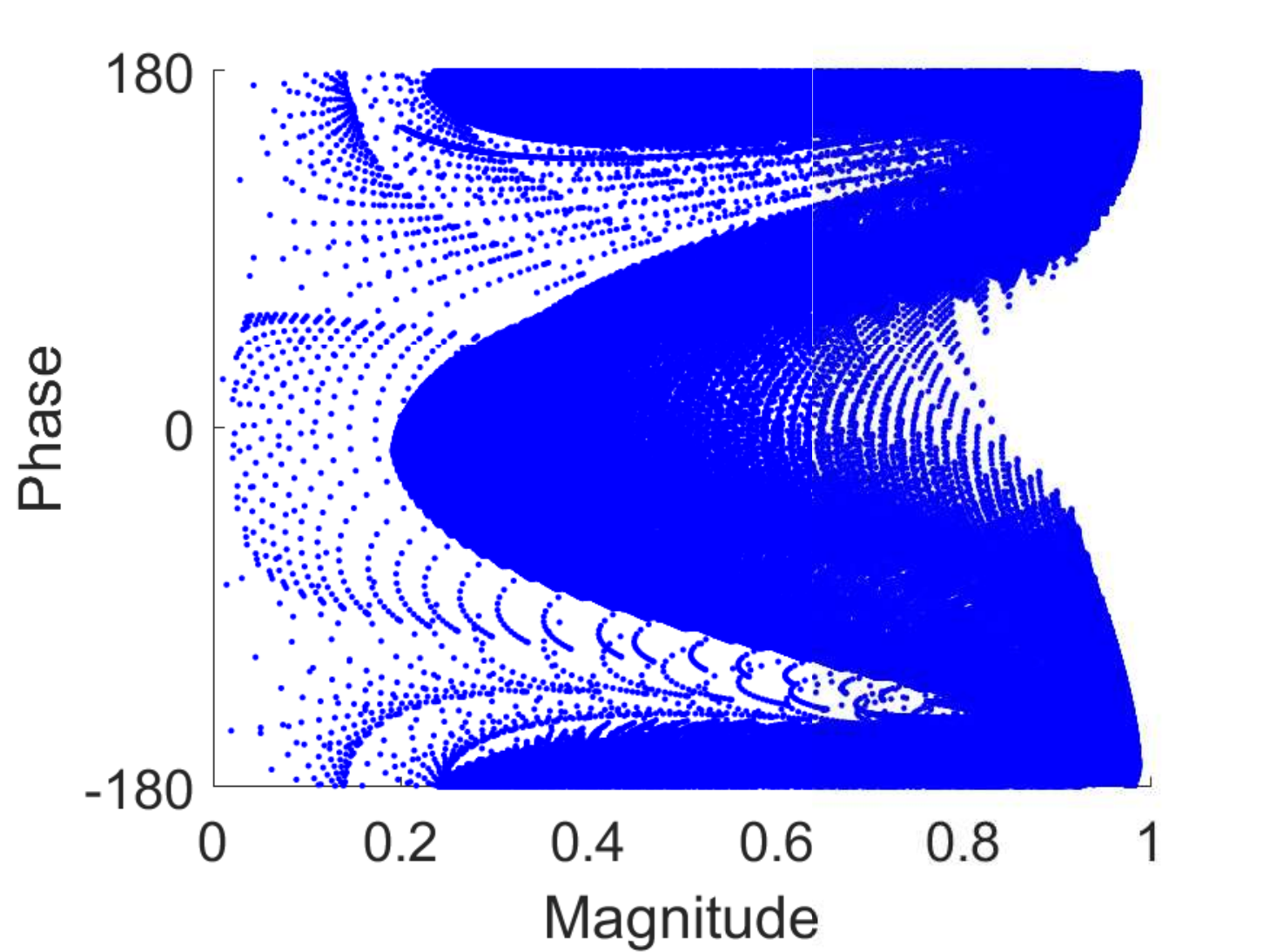}%
	}
	&
	\subfloat[]{\includegraphics[width=5.7cm]{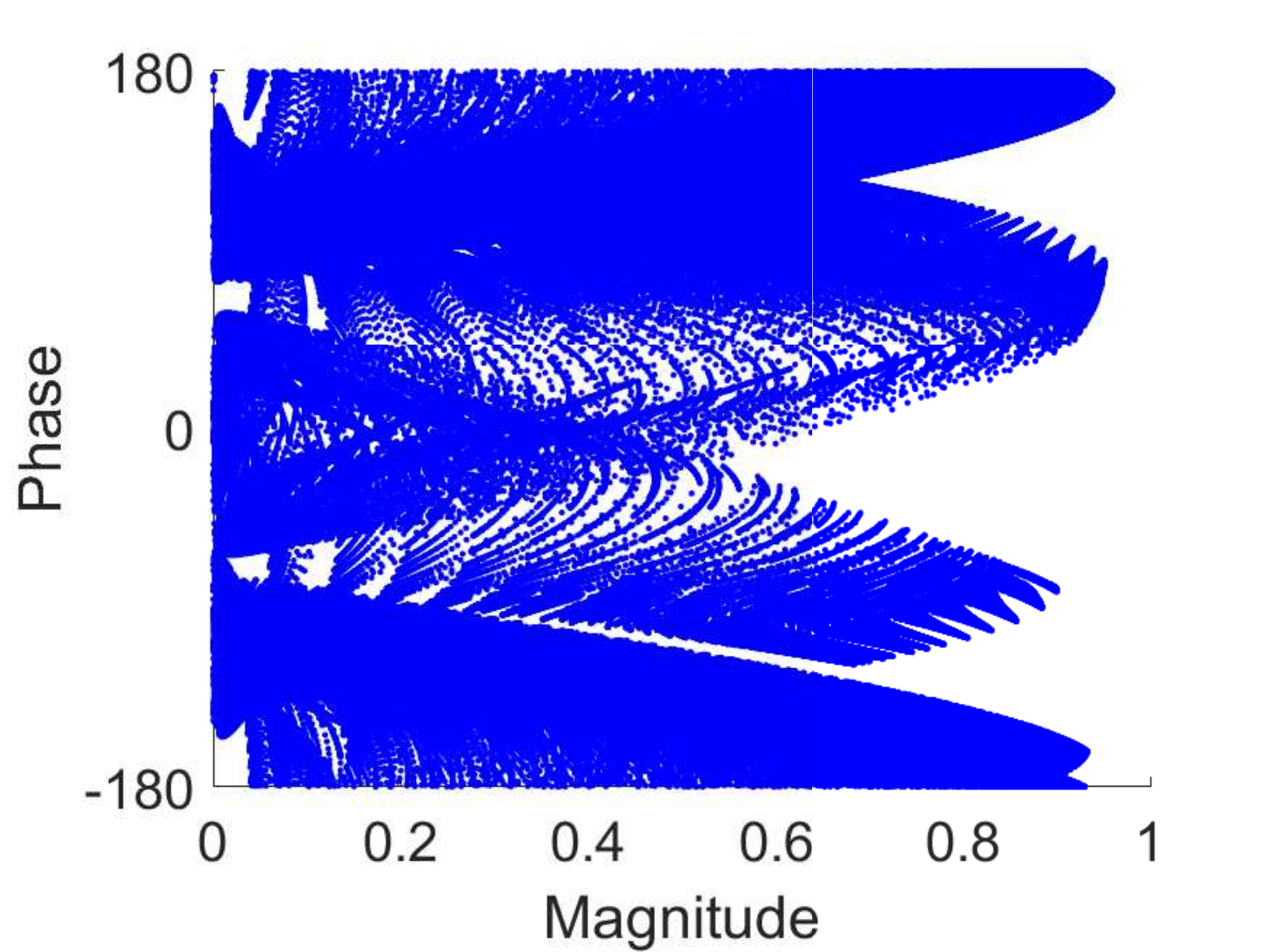}%
	}
	&
		\subfloat[]{\includegraphics[width=5.7cm]{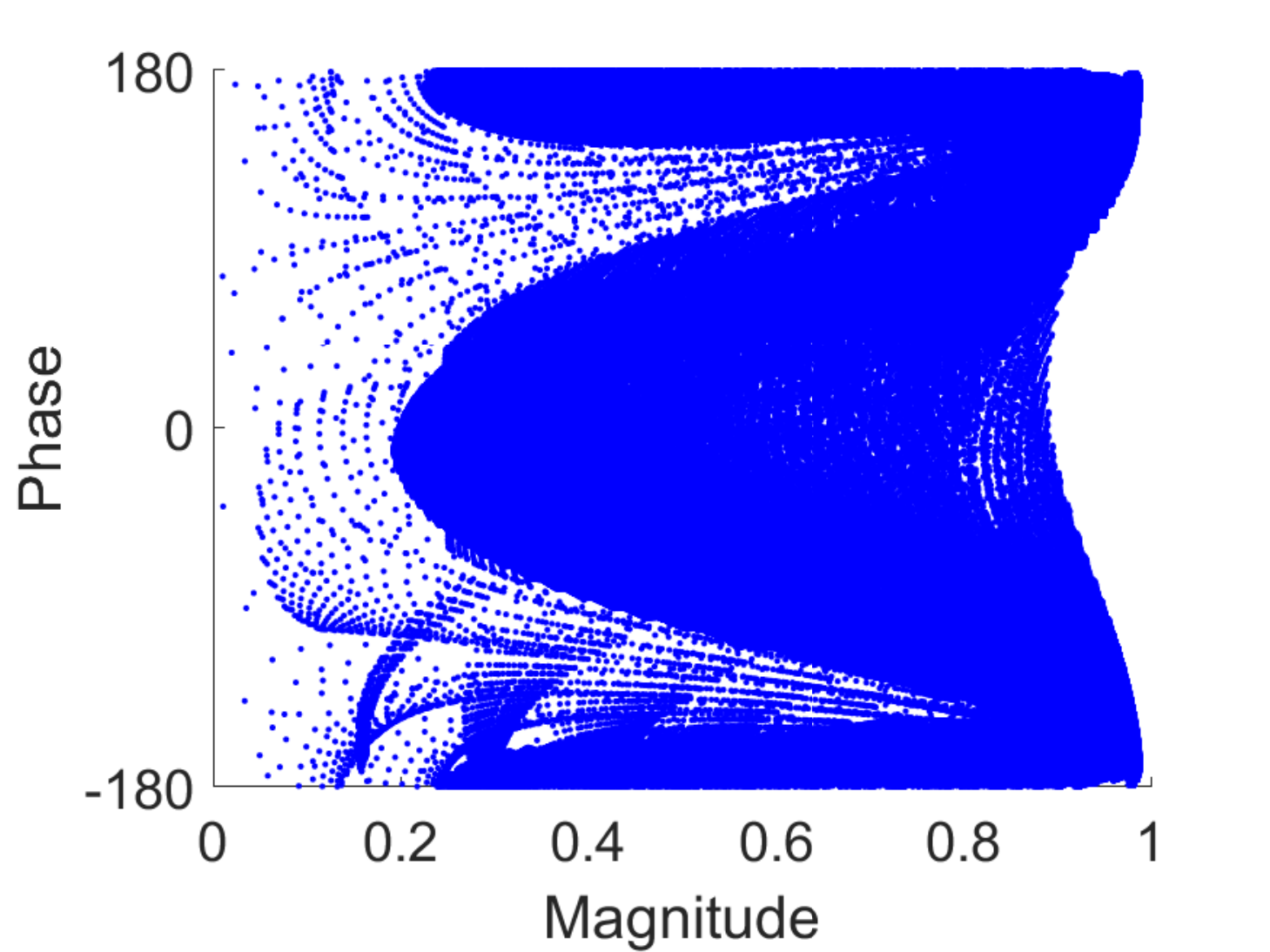}%
	}	
\end{tabular}}
\caption{S-parameter map of achievable values with 3 cascaded dog-bones when losses are limited up to $15\%$. (a) $S_{11}$, (b) $S_{12}$ and (c) $S_{22}$.}\label{fig:SParamMap3Layers}
\end{figure*}

\begin{figure*}[t]
\centering{
	\begin{tabular}{ccc}
	\subfloat[]{\includegraphics[width=5.7cm]{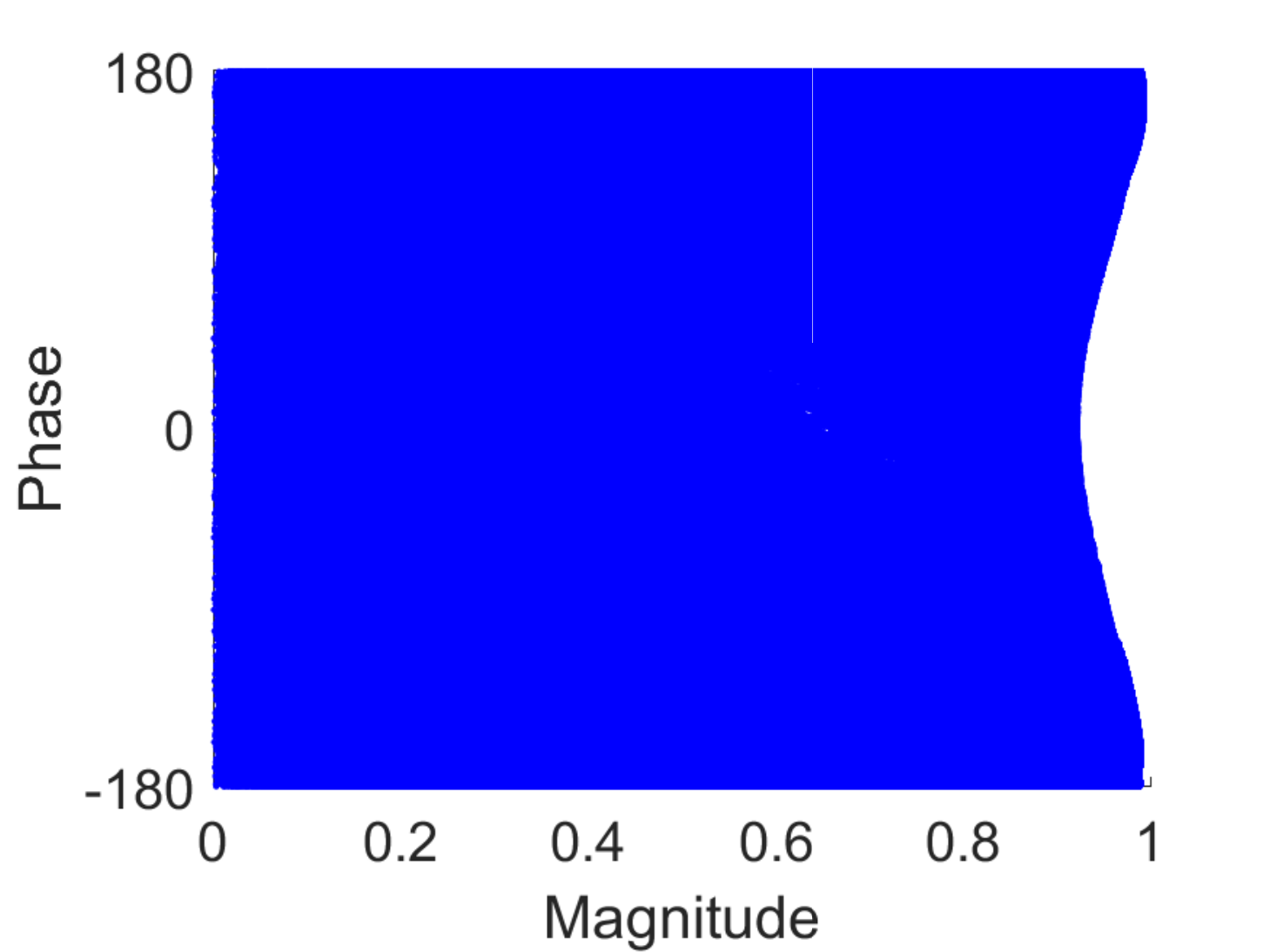}%
	}
	&
	\subfloat[]{\includegraphics[width=5.7cm]{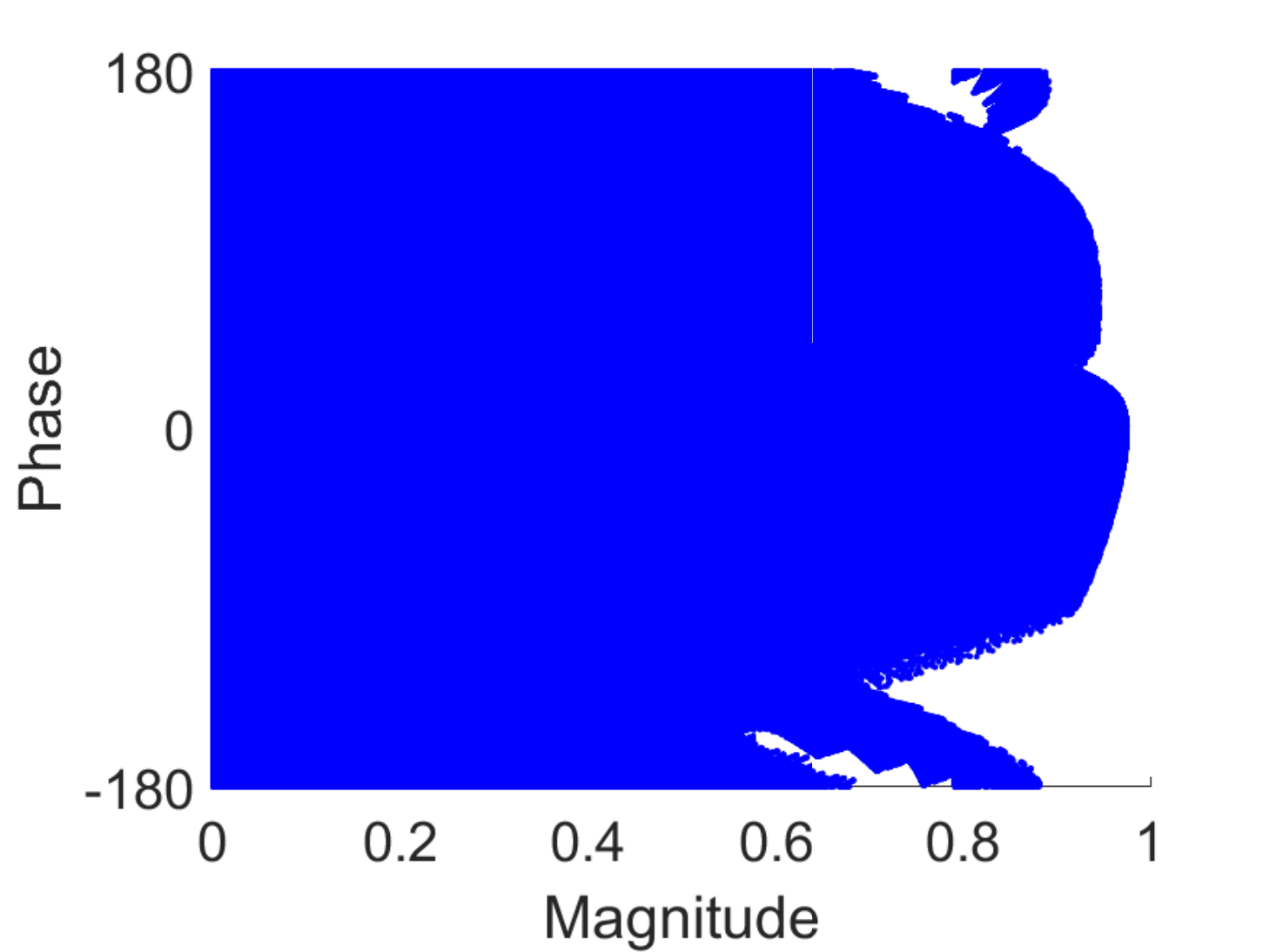}%
	}
	&
		\subfloat[]{\includegraphics[width=5.7cm]{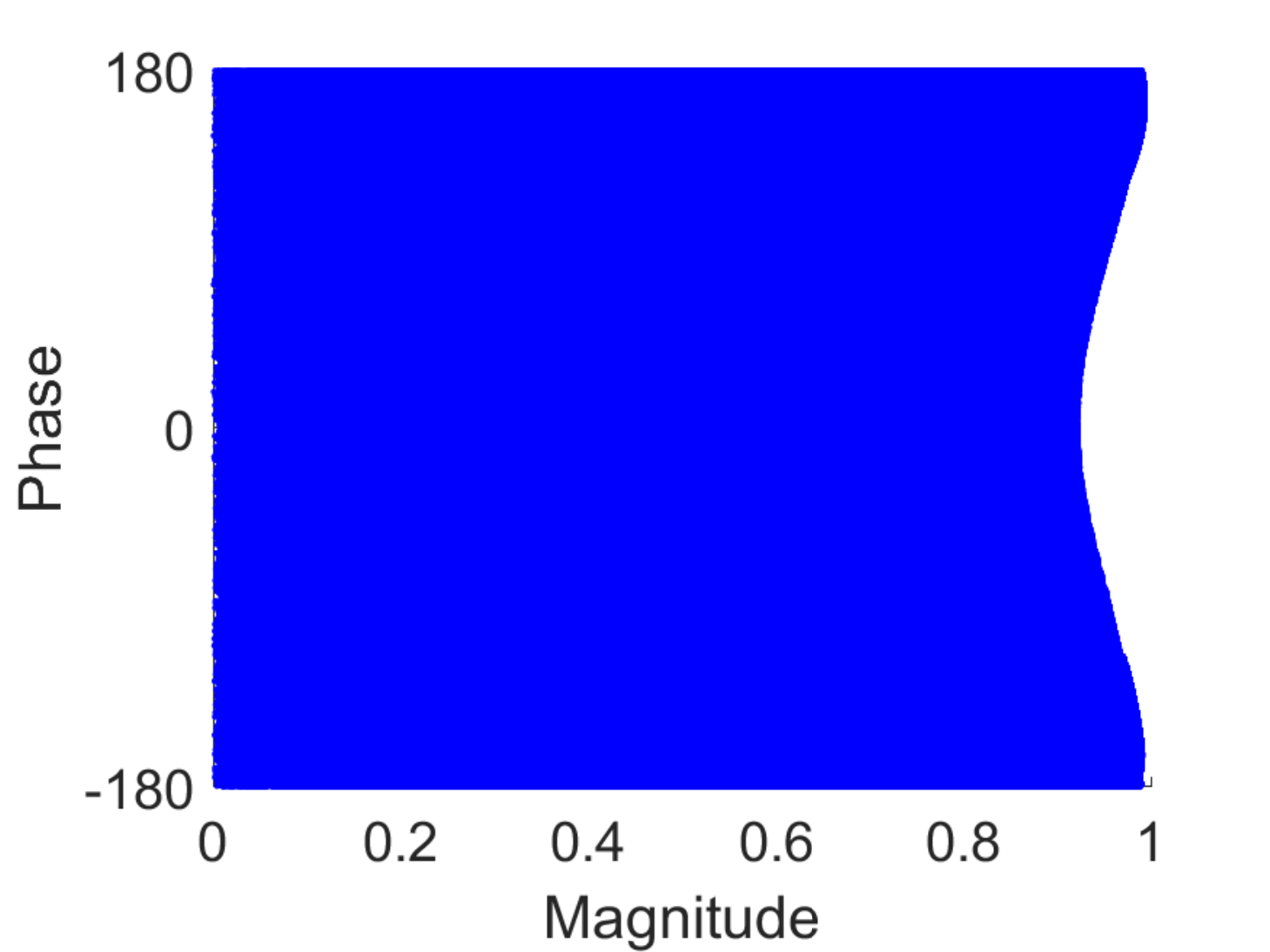}%
	}	
\end{tabular}}
\caption{S-parameter map of achievable values with 4 cascaded dog-bones when losses are limited up to $15\%$. (a) $S_{11}$, (b) $S_{12}$ and (c) $S_{22}$.}\label{fig:SParamMap4Layers}
\end{figure*}

It is customary to consider normal incidence in the design of the meta-atoms. However, when the impinging angle $\theta_{in}$ is far from $0^\circ$, the scattering properties of the unit-cell can differ significantly from the case of normal incidence, especially for thicker stack ups. Therefore, if, as previously mentioned, high $\theta_{in}$ is utilized to decrease losses, then the local design of the unit-cell should consider this phenomenon. Specifically, the equivalent circuit of the dielectrics should consider oblique incidence, since the dielectric layers are significantly thick. For this reason, when calculating the cascaded response, we have used a TL model for the dielectrics considering oblique incidence. Since $\theta_{out}$ is different from $\theta_{in}$, a question might arise about which angle to consider for the design of the unit-cells. As this design is related to the local behavior of the metasurface, we considered $\theta_{in}$ as the angle of incidence for the unit-cell design, since the resulting $\theta_{out}$ is a consequence of the entire period and cannot be related easily to a local phenomenon. So, we have applied refraction laws considering the incident angle $\theta_{in}$ and used this refracted angle in the substrate to calculate the TL parameters in the model of Fig. \ref{fig:ModelRealUnitCell} as follows:
\begin{equation}
\begin{split}
Z_{sub}(\theta_r)&=\frac{Z_{sub}(0)}{\cos(\theta_r)}\\
\Theta_{sub}(\theta_r)&=\beta_{sub}(0)\cos(\theta_r)t\,,
\end{split}
\end{equation}
where the $Z_{sub}(\theta_r)$ and $\Theta_{sub_r}(\theta_r)$ are the characteristic impedance and electric length of the TL model for the substrate when considering oblique incidence, with $\theta_r$ given by the refraction law as $\asin(\frac{1}{\sqrt{\epsilon_{sub}}}\sin(\theta_{in}))$, $\beta_{sub}$ is the phase constant, and $t$ is the substrate thickness. For instance, for $\theta_{in}=48.6^\circ$ and $\epsilon_{sub}=12.94$ (Design 1 and 2 of Section \ref{s:Designs&Simulation}), the resulting $\theta_r$ is $12.03^\circ$. Additionally, the simulation of each layer has been made taking into account $\theta_{in}$ instead of normal incidence in the periodic boundaries. However, insignificant angle dispersion has been found for the impedance values of the dog-bone of a single layer, as illustrated in Fig. \ref{fig:Xbottop}. 

Once the cascaded response of all the geometrical combinations has been obtained incorporating the individual sheet impedances (complex values) available (Fig. \ref{fig:Xbottop}) and the ABCD matrices for the middle layers into the TL model of Fig. \ref{fig:ModelRealUnitCell}, the choice of the geometrical parameters for the targeted design is done by searching for the minimum deviation between the required and available S-parameters of the unit-cell. A simple algorithm is used for this purpose, calculating separately the differences in magnitude and phase. We establish some maximum deviations for the magnitude and phase of each S-parameter and a maximum percentage of losses (calculated as $1-|S_{11}|^2-|S_{12}|^2$) above which the unit-cell geometry is discarded. Since the phases determine the pointing direction, we have introduced the use of weights to give priority to the phase rather than the magnitude. Finally, we choose the geometrical configuration that has the minimum global difference calculated as previously explained to the targeted S-parameters.

In order to illustrate the need for using four layers instead of three, Fig. \ref{fig:SParamMap3Layers} shows the achievable S-parameters of all possible combinations of three layers when the total losses are limited to $15\%$. 
Since the parametric sweep is done completely in Matlab by using the TL model, after extracting the individual sheet impedances for the bottom and top layers and ABCD matrices for the middle layers in full-wave simulations, it allows exploration of a large number of combinations in a reasonable run time. It can be observed, that there is a significant number of phase/magnitude pairs that cannot be achieved. On the contrary, as illustrated in Fig. \ref{fig:SParamMap4Layers}, if we use four layers, the range of values of magnitude and phase of the S-parameters for a limited amount of losses is much wider.

\begin{figure*}[t!]
\centering
\includegraphics[width=18cm]{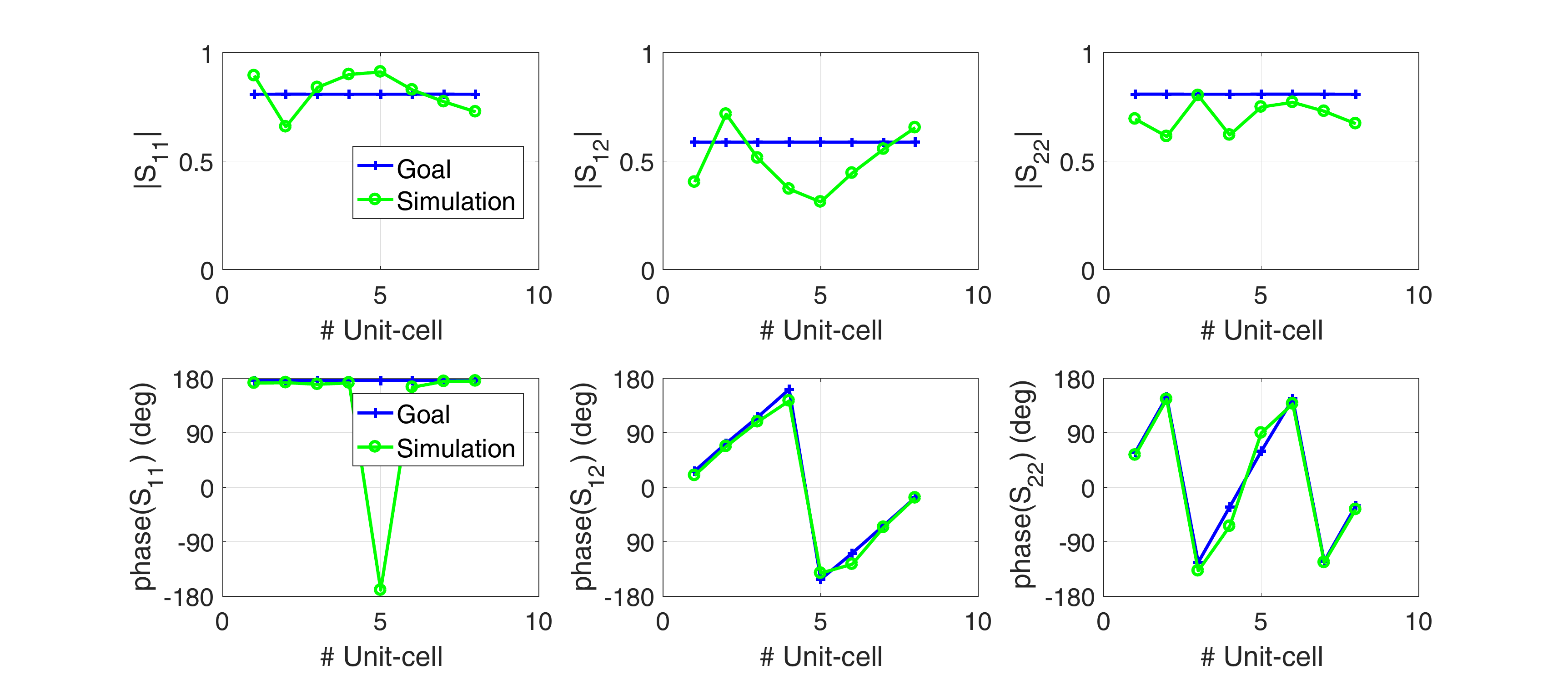}
\caption{S-parameters of each unit-cell inside the period for Design 1: targeted parameters from the theoretical derivation vs. obtained with the chosen geometry in simulations assuming local periodicity.}
\label{fig:AchievedVsGoalSParam_DesignC}
\end{figure*}

\section{Designs and Simulation Results}\label{s:Designs&Simulation}
In order to prove the concept, we have carried out three different designs. To show that the leakage factor $\alpha$ and the pointing direction $\theta_{out}$ can be arbitrarily set, we have chosen two designs with the same pointing direction and different $\alpha$ values and another one with a different pointing direction. Since the broadside case is controversial in LWAs \cite{Ol93, Li02}, we have chosen this direction for two of the three designs to additionally prove that in the proposed concept there is no open-stopband effect at broadside. The three implementations are designed to radiate around $90\,\%$ of the power, so the lengths are set according to the leakage factors. The periods have been chosen so that the propagation angle of the waveguide mode, $\theta_{in}$, is high to reduce the losses, as previously argued in Section \ref{s:PhysicalRealization}. For the three cases, the waveguide height has been arbitrarily set to $d=0.75\,\lambda_0$. Table \ref{tab:ParameterDesigns} summarizes the design parameters of the three designs. It could be mentioned that the beam pointing angle of the non-broadside design has arbitrarily been chosen to be backwards; however, a design for a forward angle would not entail any additional challenges (a design example of an extreme forward angle can be found in \cite{Ab17}).

As can be seen in Table \ref{tab:ParameterDesigns}, the periods are rather large, so according to the Floquet's theorem, for each of the designs, there would be more spatial harmonics able to radiate. For instance, for Designs 1 and 2, the fundamental spatial harmonic ($\theta=48.6^\circ$) and the harmonic $-2$ ($\theta=-48.6^\circ$) would be within the radiation cone. However, it is expected, as previously argued, that these modes will not be excited and, then, a single beam, corresponding to $m=-1$, will appear in the radiation pattern.

\begin{figure}[t!]
\centering
\includegraphics[width=\columnwidth]{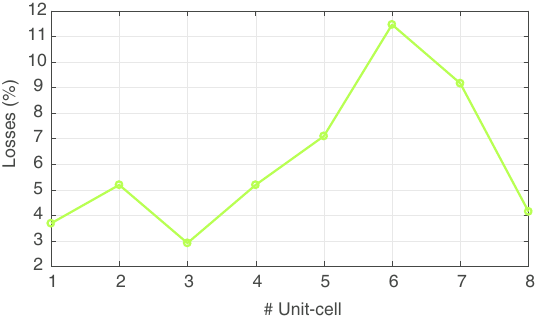}
\caption{Simulated total losses of each unit-cell inside the period  of Design 1 assuming local periodicity.}
\label{fig:AchievedLossesUnitCells_DesignC}
\end{figure}

As an example, Fig. \ref{fig:AchievedVsGoalSParam_DesignC} shows the comparison of the targeted S-parameters of the unit-cells of Design 1 (broadside, $\alpha=0.013k_0$) derived from theory with those obtained from simulation of the physical (printed-circuit-board-compatible) realization of the dog-bones (assuming local periodicity) after designing them following the aforementioned synthesis procedure. One should notice that this parameter specification indicates that bianisotropic meta-atoms are required to implement the desirable LWA, as the reflection phase from the bottom ($S_{11}$) and top ($S_{22}$) faces of the metasurface should be different. This is crucial to achieve the ``perfect'' performance with no spurious modes. Since the theoretical values correspond to a lossless two-port, it is not possible to obtain the exact values of the S-parameters (especially, the magnitudes). Fig. \ref{fig:AchievedLossesUnitCells_DesignC} shows the achieved losses for each unit-cell assuming local periodicity, which were limited to $15\,\%$ in the synthesis procedure.

\begin{table}
\centering
	\caption{Summary of the design parameters for the three design examples.}
	\label{tab:ParameterDesigns}
\begin{tabular}{|l||c|c|c|}
\hline
	Parameter & Design 1 & Design 2 & Design 3 \\
	\hline
	$\theta_{out}$ [$^\circ$]	& 0 & 0 & -20\\
	$\theta_{in}$ [$^\circ$] & 48.6 & 48.6 & 41.1\\
	$p$ [$\lambda_0$] & 8/6 & 8/6 & 1\\
	$L$ [$\lambda_0$] & 15 & 10 & 10\\
	$\alpha$ [$k_0$] & 0.013 & 0.02 & 0.02\\
	$d$ [$\lambda_0$] & 0.75 & 0.75 & 0.75\\
\hline
\end{tabular}
\end{table}

Tables \ref{tab:GeometryDesign1}, \ref{tab:GeometryDesign2} and \ref{tab:GeometryDesign3} show the resulting lengths of the dog-bones for each unit-cell of each design, to be used in the final layout of the metasurfaces.

\begin{table}[h]
	\caption{Geometrical parameters of Design 1}
\setlength\tabcolsep{4pt}
\begin{tabular}{l|*8{c}}
	\#Cell & 1 & 2 & 3 & 4 & 5 & 6 & 7 & 8 \\
	\hline
	$L_{bot}$ [mil]	& 25 & 18 & 54 & 24.5 & 14 & 25.5 & 19.5 & 9\\
	$L_{mid1}$ [mil] & 9 & 16 & 5.5 & 15 & 54.5 & 54.5 & 16.5 & 5\\
	$L_{mid2}$ [mil] & 40.5 & 54.5 & 54.5 & 54.5 & 46.5 & 49.5 & 41.5 & 40.5\\
	$L_{top}$ [mil] & 9.5 & 55 & 34.5 & 31 & 31 & 28 & 13 & 9\\
\end{tabular}
	\label{tab:GeometryDesign1}
\end{table}

\begin{table}[h]
	\caption{Geometrical parameters of Design 2}
\setlength\tabcolsep{4pt}
\begin{tabular}{l|*8{c}}
	\#Cell & 1 & 2 & 3 & 4 & 5 & 6 & 7 & 8 \\
	\hline
	$L_{bot}$ [mil]	& 55 & 50.5 & 26.5 & 27.5 & 33 & 20 & 17.5 & 10.5\\
	$L_{mid1}$ [mil] & 5 & 54.5 & 52.5 & 5.5 & 54.5 & 36 & 30.5 & 5\\
	$L_{mid2}$ [mil] & 32 & 30.5 & 9 & 51.5 & 49 & 38.5 & 40 & 40.5\\
	$L_{top}$ [mil] & 10.5 & 11.5 & 12 & 31 & 31 & 18 & 15 & 9\\
\end{tabular}
	\label{tab:GeometryDesign2}
\end{table}

\begin{table}[h]
	\caption{Geometrical parameters of Design 3}
\begin{tabular}{l|*6{c}}
	\#Cell & 1 & 2 & 3 & 4 & 5 & 6 \\
	\hline
	$L_{bot}$ [mil]	& 5 & 5 & 17 & 15.5 & 16.5 & 5\\
	$L_{mid1}$ [mil] & 47 & 48 & 42 & 43 & 30 & 47.5\\
	$L_{mid2}$ [mil] & 27.5 & 54.5 & 41 & 32.5 & 10.5 & 30.5\\
	$L_{top}$ [mil] & 15.5 & 27.5 & 11.5 & 17 & 17.5 & 17.5\\
\end{tabular}

	\label{tab:GeometryDesign3}
\end{table}

\begin{figure}[thb!]
\centering
\includegraphics[width=\columnwidth]{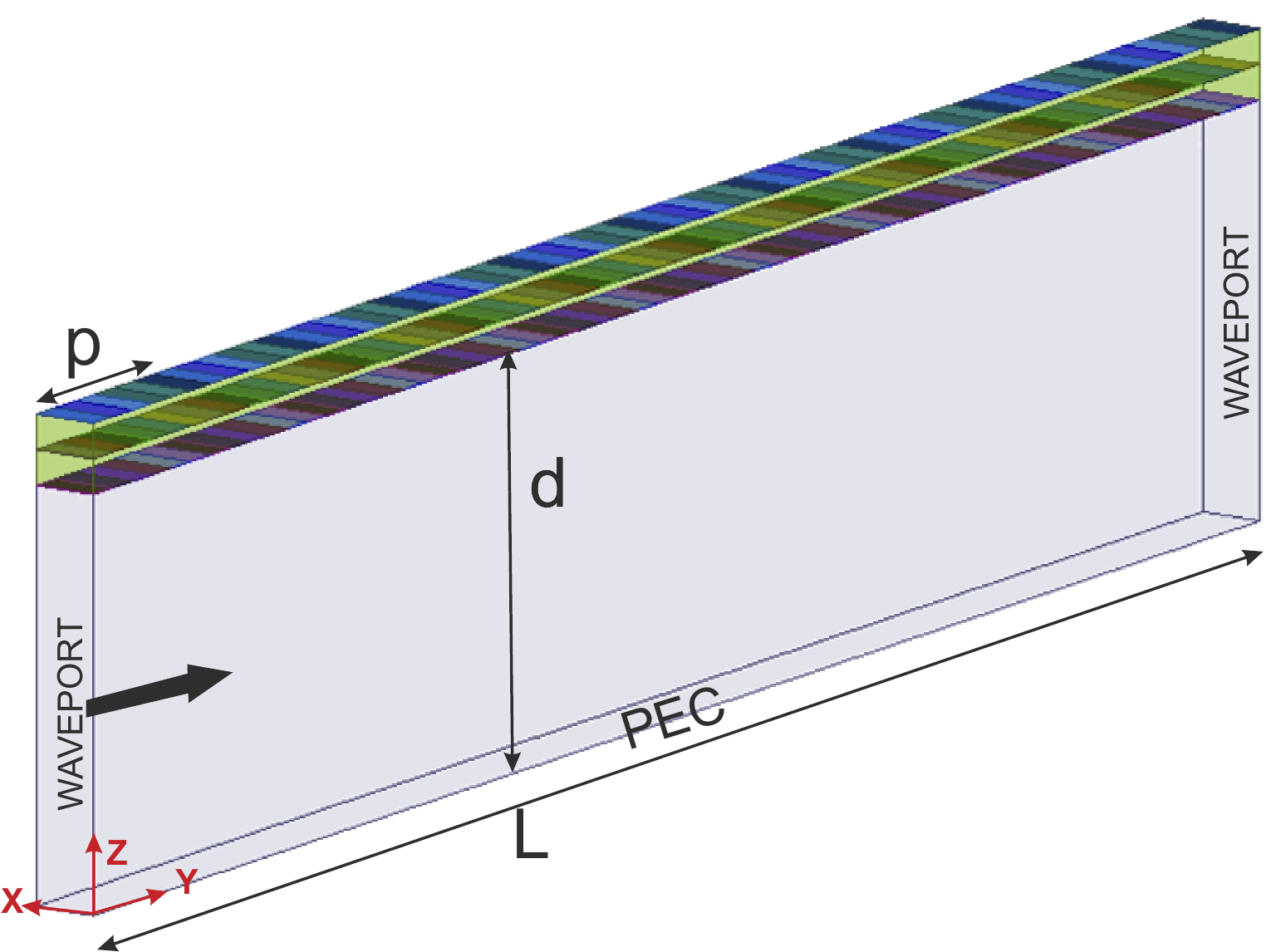}
\caption{Schematic of simulation when using impedance sheets for the metasurface (`idealized simulation').}
\label{fig:SchematicLWASim}
\end{figure}

In order to prove the concept, two different types of simulations have been carried out to be compared with the theoretical performance (\ref{eq:FieldBelow})-(\ref{s:ConditionEout}), both with the electromagnetic simulator HFSS. In a first stage, we want to verify that the macroscopic design (\ref{eq:MetaParam}) is correct, independently of the geometry used later on for the physical implementation of the metasurface (microscopic design). For this purpose, impedance boundary conditions have been used in HFSS in order to mimic the reactance sheets needed to implement the metasurface according to the design synthesis shown in Fig. \ref{fig:CascadedSheets}. The simulation schematic for this case can be observed in Fig. \ref{fig:SchematicLWASim}. We have used a small substrate thickness (2\,mil) for this idealized simulation, and the impedance sheets used were completely lossless. 

The other type of simulation is the one in which the microscopic design for the metasurface is taken into account, so the unit-cells are physically implemented with the dog-bones as prescribed by Tables \ref{tab:GeometryDesign1}, \ref{tab:GeometryDesign2} and \ref{tab:GeometryDesign3}; copper and dielectric losses are considered in this realistic simulation. In both cases, PEC walls are used as periodic boundaries taking advantage of the TE configuration.

\begin{figure*}[!thb]
\centering{
	\begin{tabular}{ccc}
	\subfloat[]{\includegraphics[width=5.7cm]{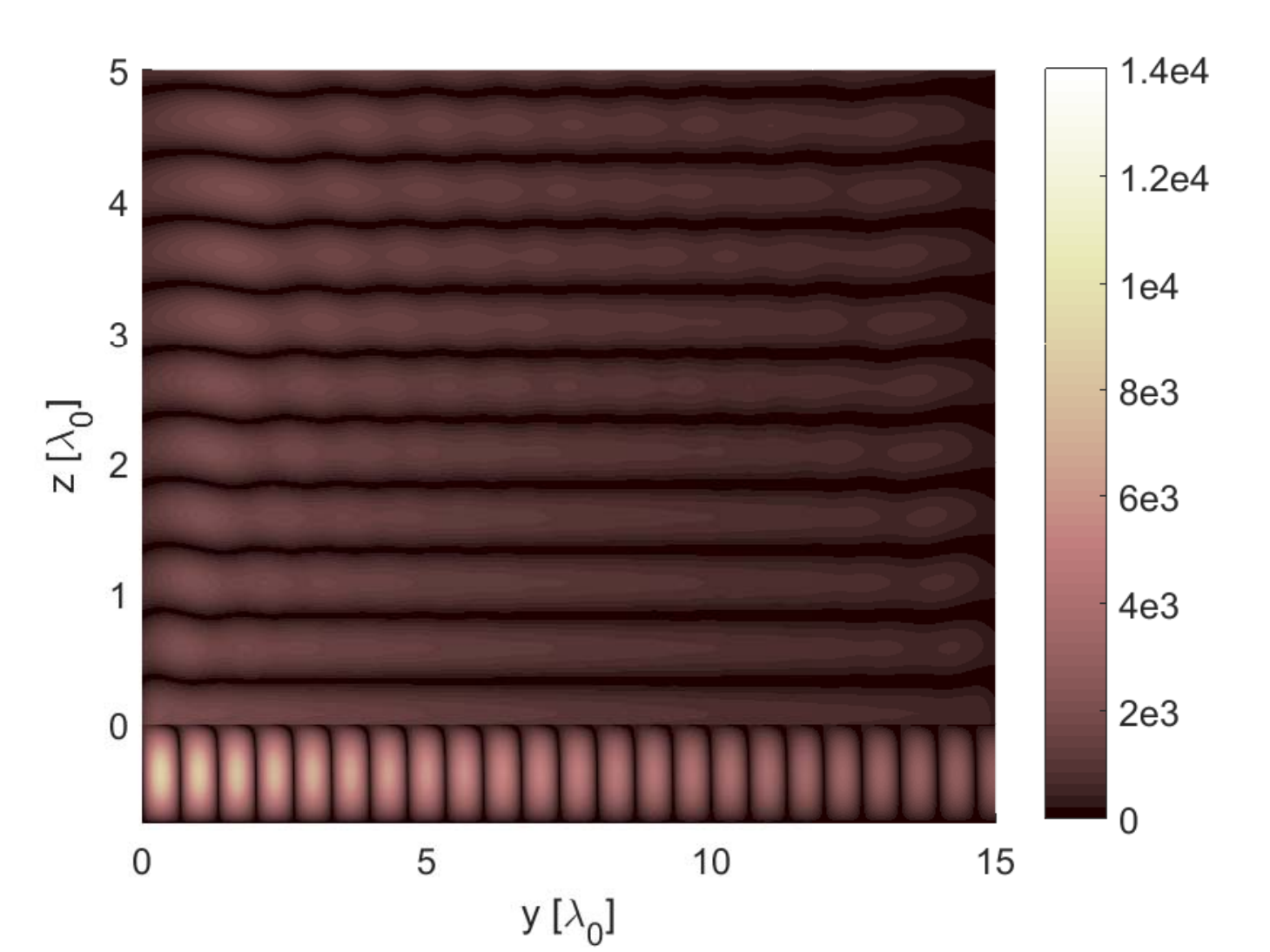}%
	}
	&
	\subfloat[]{\includegraphics[width=5.7cm]{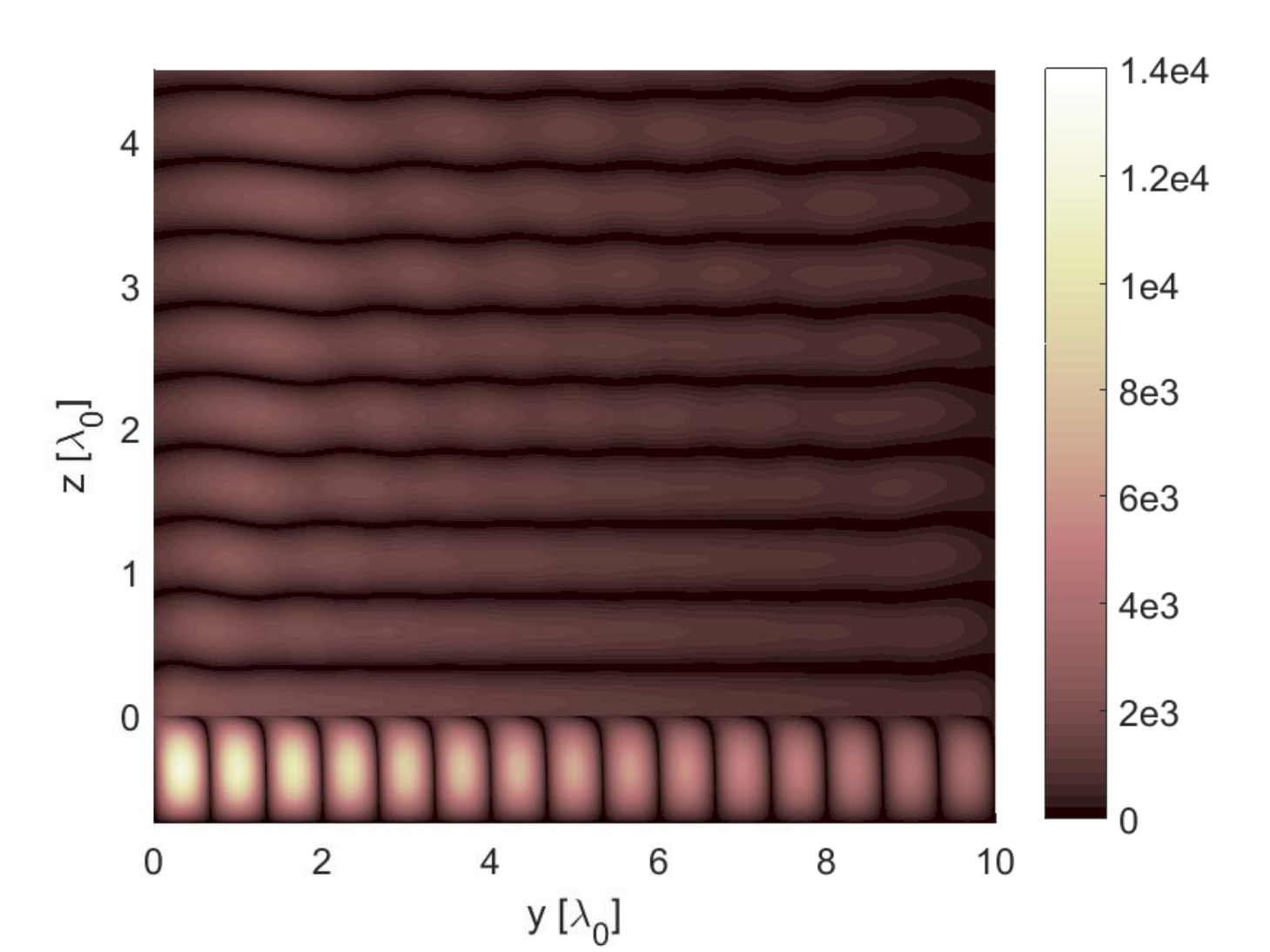}%
	}
	&
		\subfloat[]{\includegraphics[width=5.7cm]{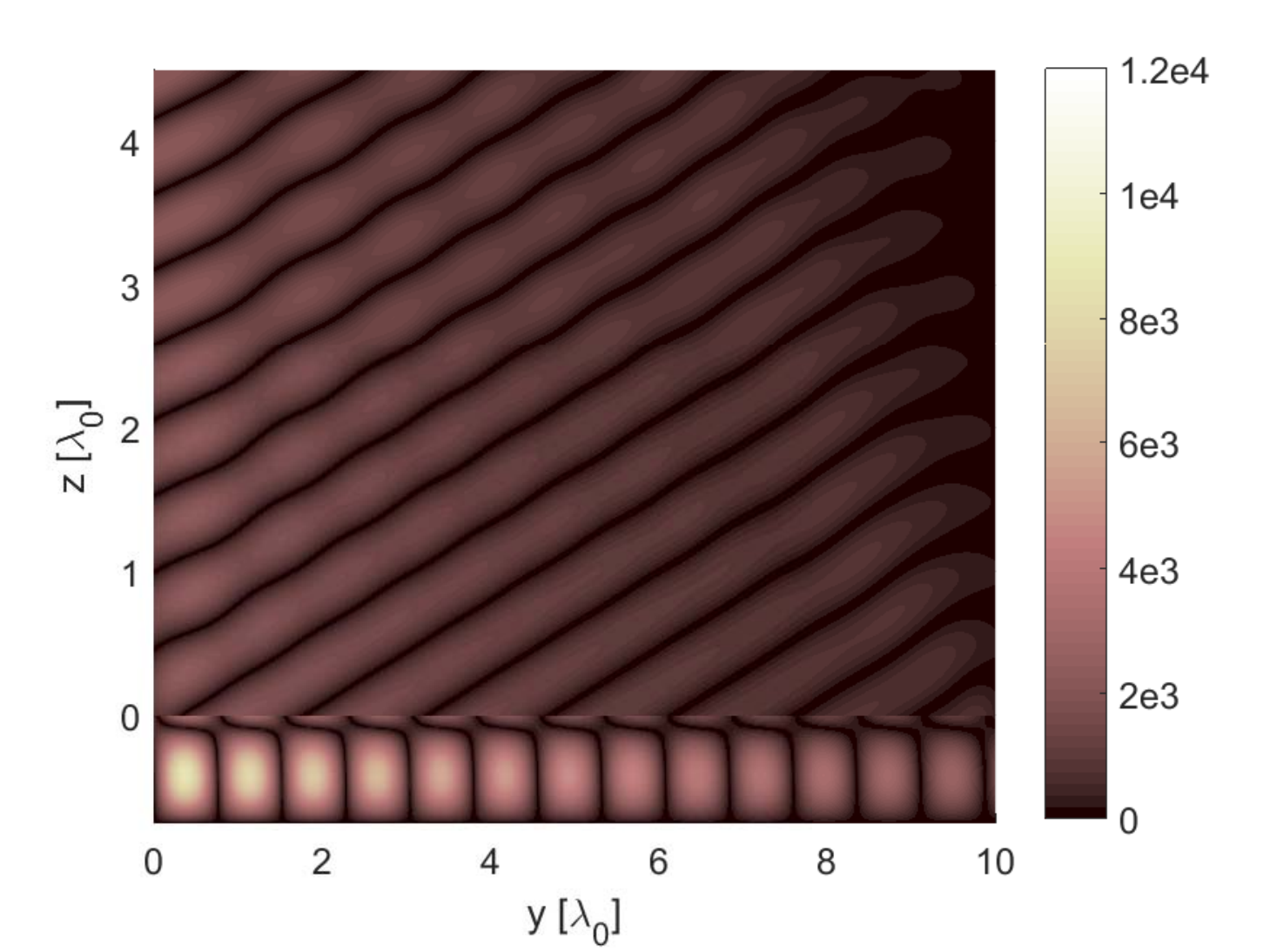}%
	}
	\\
	\subfloat[]{\includegraphics[width=5.7cm]{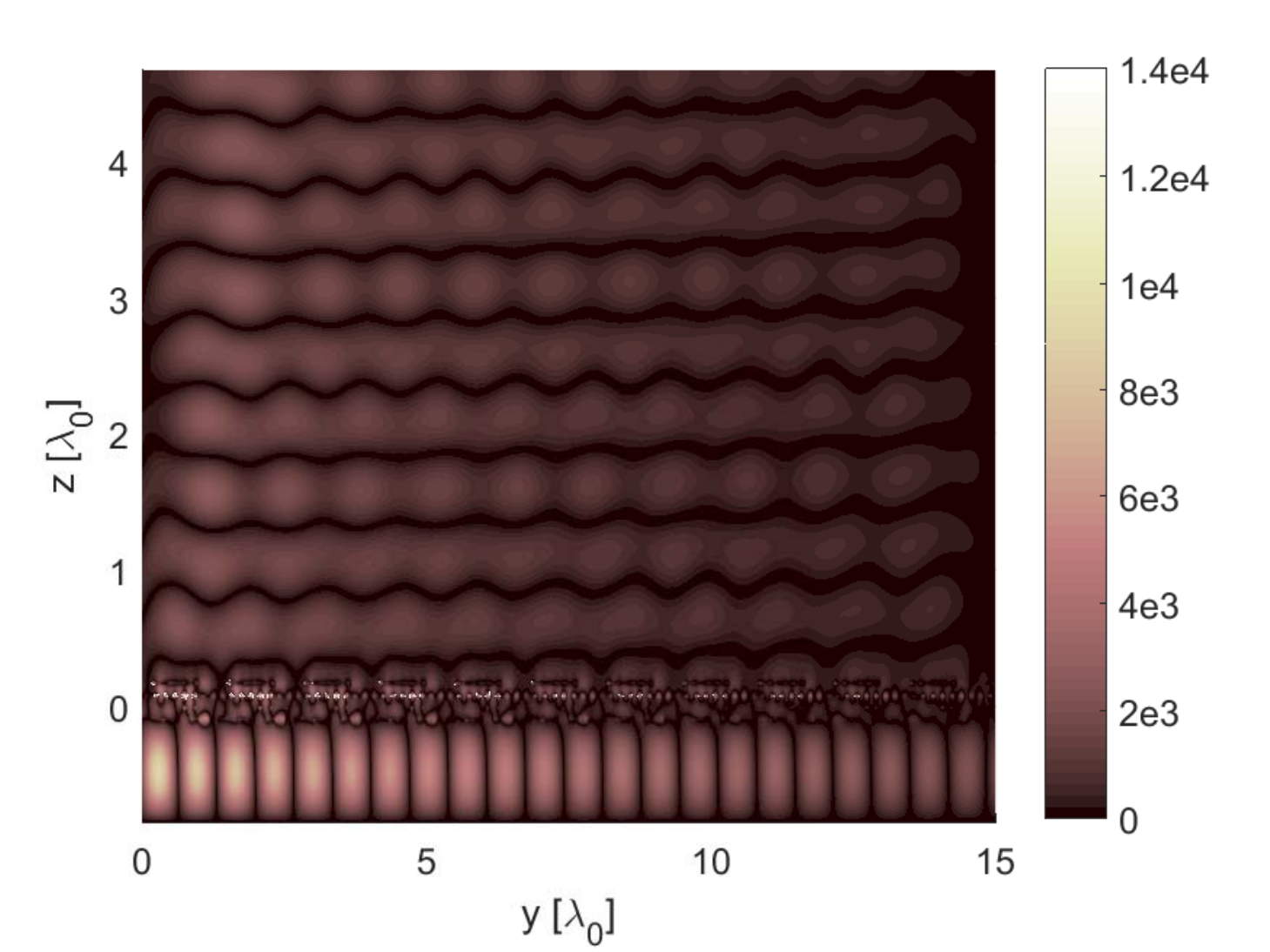}
	}
	&
	\subfloat[]{\includegraphics[width=5.7cm]{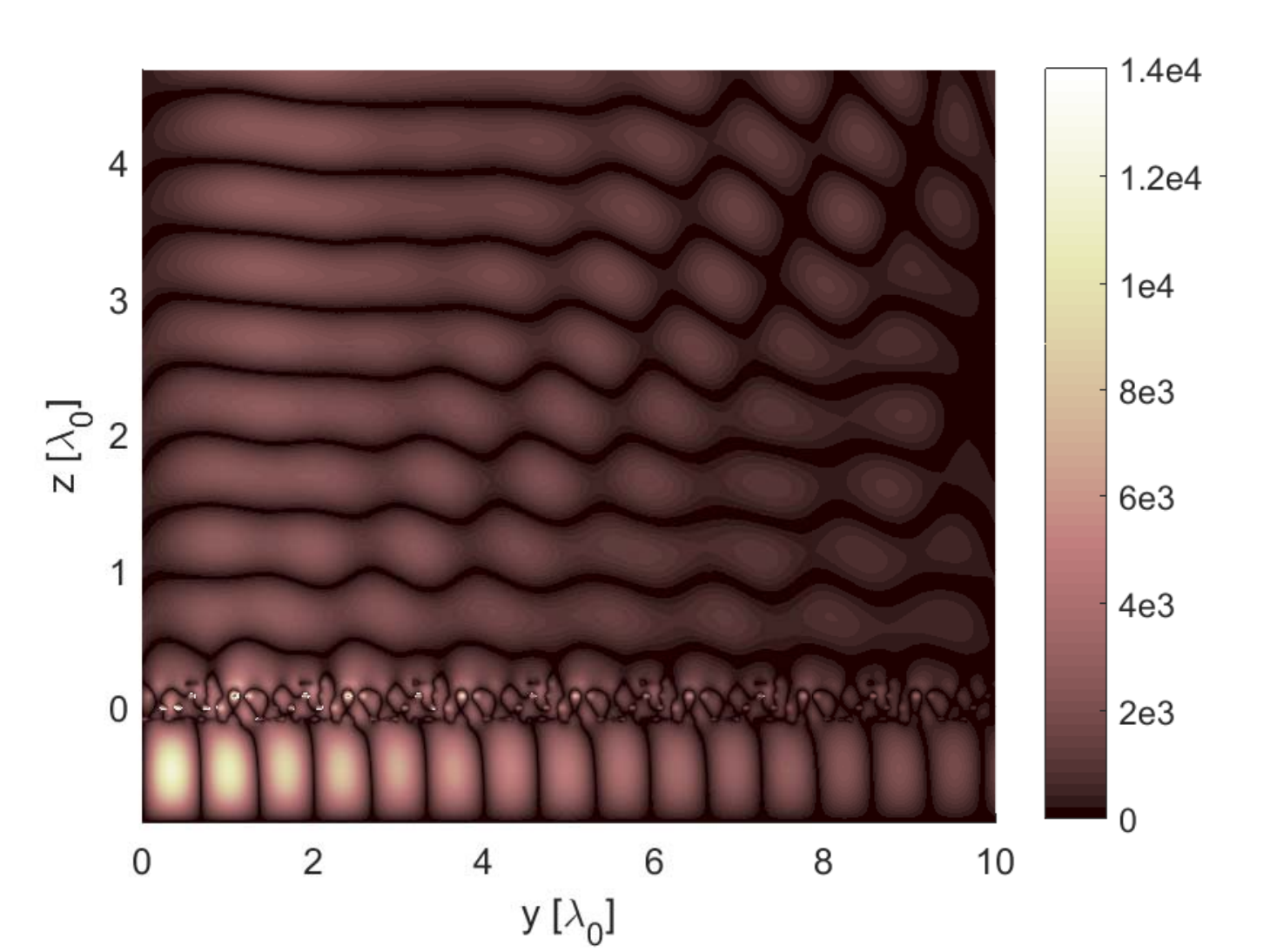}
	}
	&
	\subfloat[]{\includegraphics[width=5.7cm]{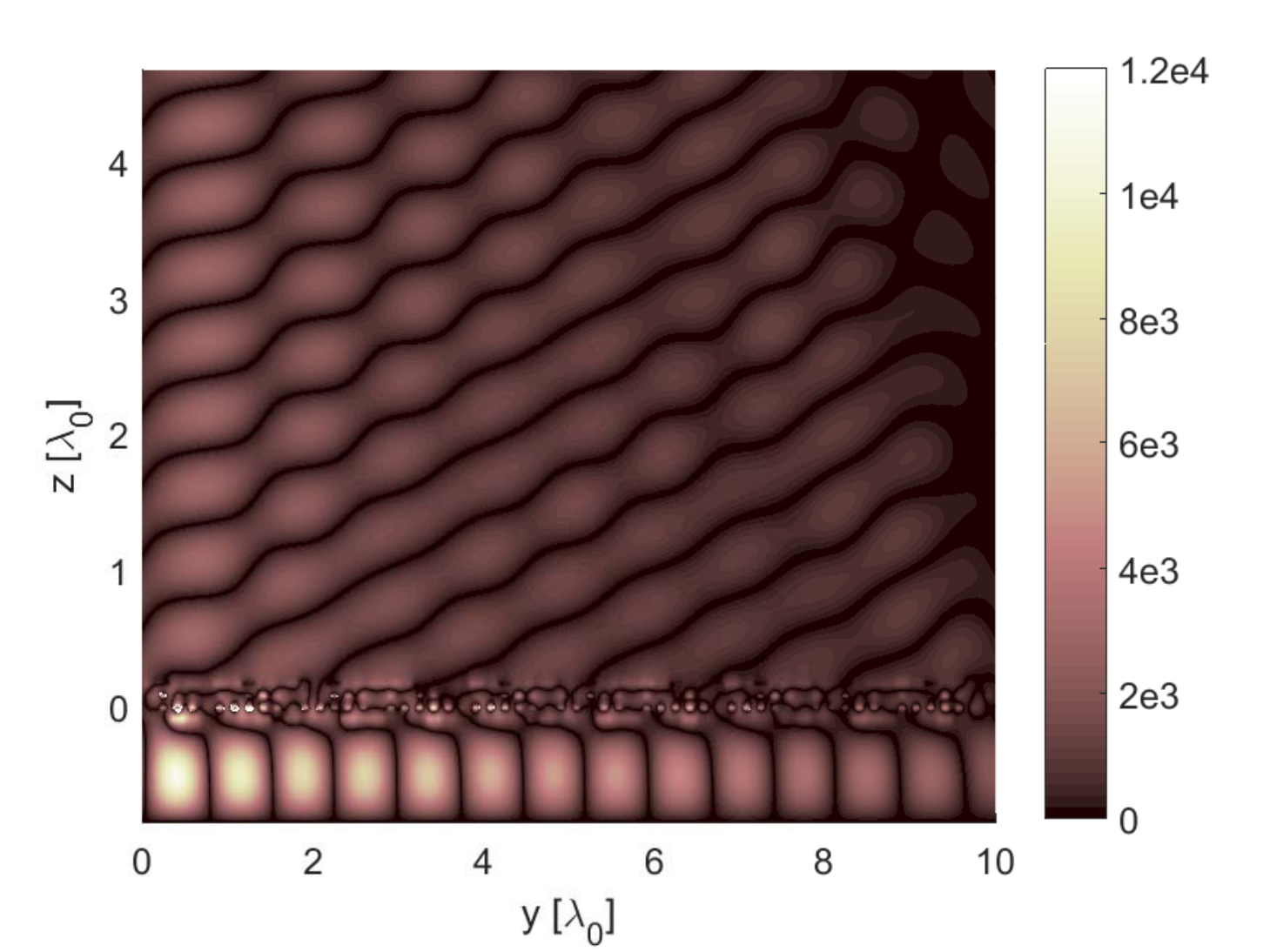}
	}		
\end{tabular}}
\caption{Field distributions $|Re(E_x(y,z)|$ (V/m) for input power of 1W over a waveport area $\lambda_0$/6 x $\lambda_0$/9.5 x $0.75\lambda_0$: (a)-(c) theoretical prediction and (d)-(f) full-wave realistic simulation for Design 1 (a and d), Design 2 (b and e) and Design 3 (c and f).}\label{fig:FieldPlots}
\end{figure*}

Fig. \ref{fig:FieldPlots} shows the plot of the magnitude of the electric field along the zy-plane, obtained from the analytical derivation and from the realistic simulation for the three design examples. Excellent agreement between both field distributions for the three examples can be observed. As designed, Designs 1 and 2 have the same input and output angles (in fact at broadside) but differ in the leakage factor, which is noticeable in the different rates of the field attenuation along the waveguide but the same mode. In contrast, Design 3 has different field patterns both inside and out of the waveguide, as expected from Table \ref{tab:ParameterDesigns}. Unlike the analytical solution, the physical implementation of the metasurface has a non-negligible thickness which leads to field interactions between the different layers forming the metasurface, observable in the field snapshots. 

\begin{figure*}[!thb]
\centering{
	\begin{tabular}{ccc}
	\subfloat[]{\includegraphics[width=5.7cm]{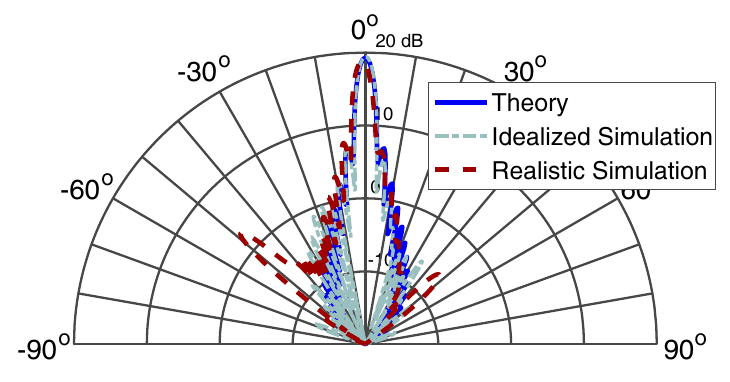}%
	}
	&
	\subfloat[]{\includegraphics[width=5.7cm]{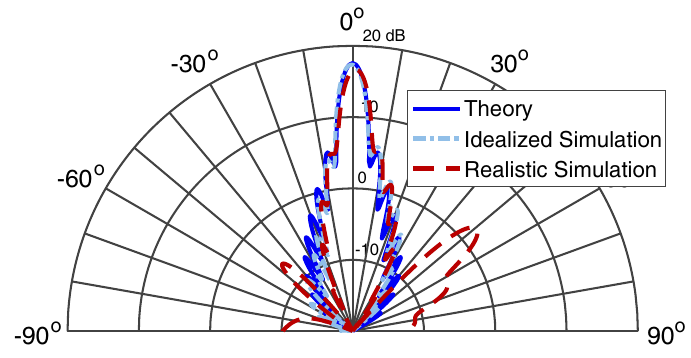}%
	}
	&
		\subfloat[]{\includegraphics[width=5.7cm]{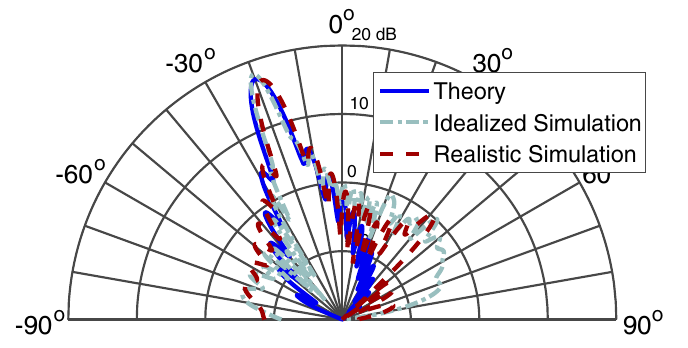}%
	}
\end{tabular}}
\caption{2D directivity radiation patterns for the yz-plane obtained from the theoretical derivation, full-wave simulation with reactance sheets(`idealized simulation') and full-wave simulation with the metasurface implemented with dog-bones. (a) Design 1, (b) Design 2 and (c) Design 3.}\label{fig:DirRadPat}
\end{figure*}

Fig. \ref{fig:DirRadPat} plots the directivity radiation patterns of the three designs. Since the proposed antenna is only directive in the zy-plane, we represent the 2D directivity as

\begin{equation}
D_{2D}=\frac{2\pi U\left(\theta, \phi=\pi/2\right)}{\int^{\pi}_{0}U\left(\theta, \phi=\pi/2\right)d\theta},
\end{equation}
where $U(\theta, \phi)$ stands for the radiation intensity.

The excellent agreement between the idealized simulation and the theoretical prediction proves the concept and shows the discretization of the continuous metasurface parameters has a negligible effect. In addition, the good agreement with the realistic simulation demonstrates the viability to implement the proposed concept using (realistically lossy) dog-bones on a commercially available substrate. The differences between the ideal metasurface and the actual implementation lead to the appearance of small side lobes corresponding to the fundamental and the $-2$ spatial harmonics of the waveguide mode (which in the ideal case are completely suppressed) and a small shift of the pointing direction. Since, in the idealized simulation, there is no noticeable deterioration with respect to the expected radiation pattern from theory, the effect of the discretization of the ideal continuous metasurface does not seem to be an issue. However, the ideal behavior is certainly affected by the discrepancies between the synthesized-unit-cell S-parameters and the targeted ones when using realistic unit-cells, since these make the theoretical boundary condition not to be perfectly fulfilled. To the extent that unit-cells with a response closer to the ideal one can be implemented (by using other substrates, geometries, etc.), these deviations can be reduced. Regardless, even with the found discrepancies, the overall antenna performance is well reproduced.

\begin{figure}[thb!]
\centering
\includegraphics[width=\columnwidth]{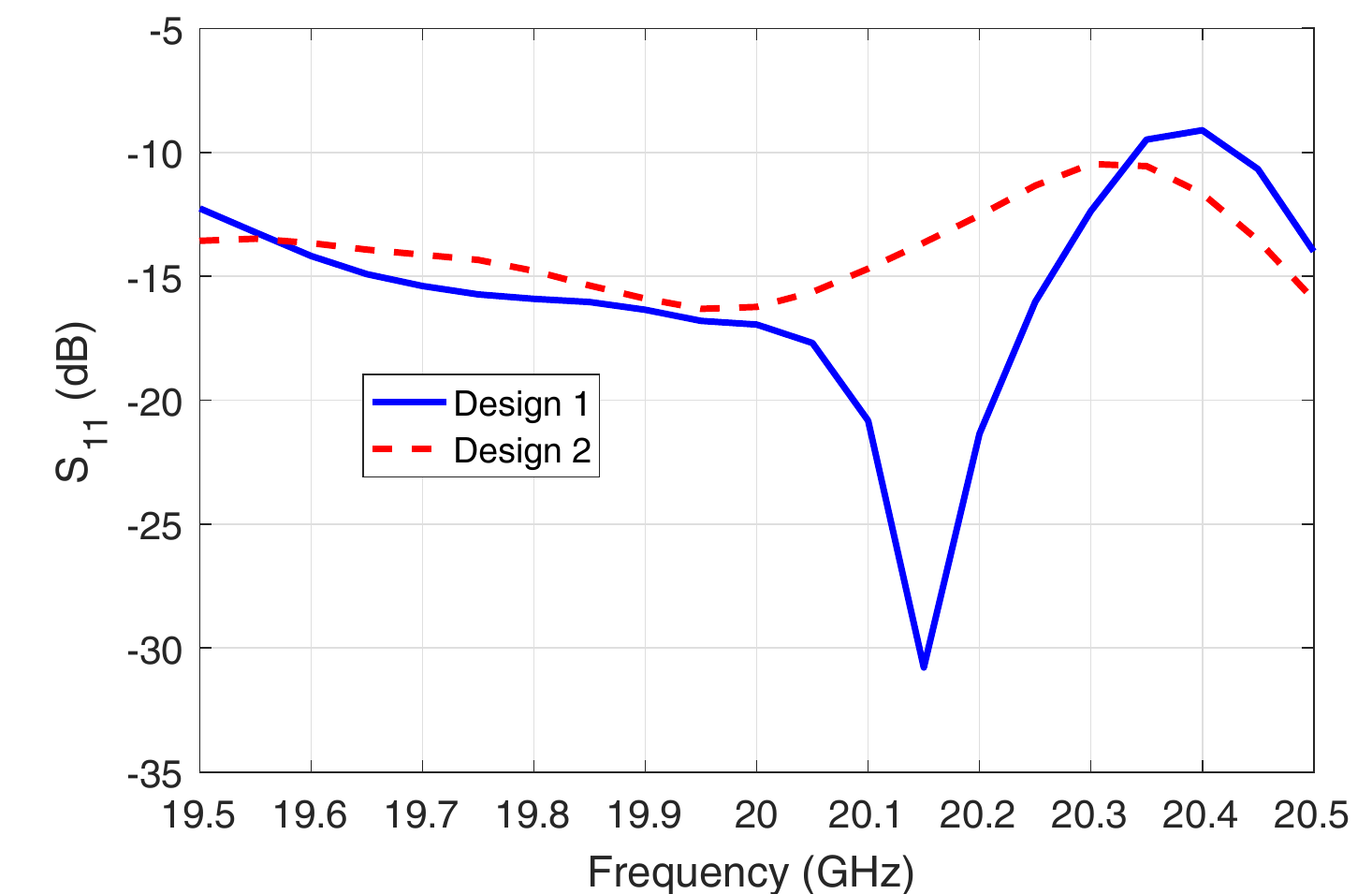}
\caption{Magnitude of the $S_{11}$ obtained from full-wave simulation with the metasurface implemented with dog-bones of Design 1 and 2.}
\label{fig:S11_DesignA&C}
\end{figure}

In order to show that broadside radiation can be achieved avoiding the open-stopband effect, Fig. \ref{fig:S11_DesignA&C} shows the reflection coefficient from the full-wave simulations of the structure with the metasurface implemented with dog-bones for Design 1 and 2. It can be observed that there is neither mismatching nor strong frequency variations around 20\,GHz (design frequency for broadside). The behavior is rather smooth. Therefore, the antenna is able to radiate at broadside without any degradation.

Finally, Table \ref{tab:PerformanceSummary} summarizes the performance obtained from theory and the two types of simulations for the three designs.

\begin{table}
	\caption{Summary of the parameter performance of the three designs.}
\setlength\tabcolsep{4pt}
\begin{tabular}{l|*6{c}}
	& $\theta_{out}$ & $D_{max}$ & $\eta_{ap}$ & $\eta_{rad}$ & SLL & $Losses$  \\
	 Design & $(^\circ)$ & (dBi) & (\%) & (\%) & (dBi) & (\%)  \\
	\hline
	Design 1 (Theory) & 0 & 19.3 & 90.6 & 91.3 & - & - \\
	Design 1 (Ideal. Sim.) & 0.1 & 19.4 & 92.9 & 86.2 & - & - \\
	Design 1 (Real. Sim.) & -0.7 & 18.7 & 79.3 & 66.4 & -15.7 & 29.5 \\
	\hline
	Design 2 (Theory) & 0 & 17.6 & 90.8 & 91.9 & - & - \\
	Design 2 (Ideal. Sim.) & -0.2 & 17.3 & 93.8 & 93.8 & - & - \\
	Design 2 (Real. Sim.) & 0.6 & 16.7 & 74.3 & 78.5 & -14.3 & 18.3 \\
	\hline
	Design 3 (Theory) & -20 & 17.3 & 85.5 & 91.9 & - & -\\
	Design 3 (Ideal. Sim.) & -20 & 18.1 & 68.0 & 94.5 & - & - \\
	Design 3 (Real. Sim.) & -18.6 & 16.8 & 76.6 & 68.0 & -16.5 & 28.3 \\
\end{tabular}
	\label{tab:PerformanceSummary}
\end{table}

\section{Experimental Validation}
\begin{figure}[t!]
\centering
\includegraphics[width=\columnwidth]{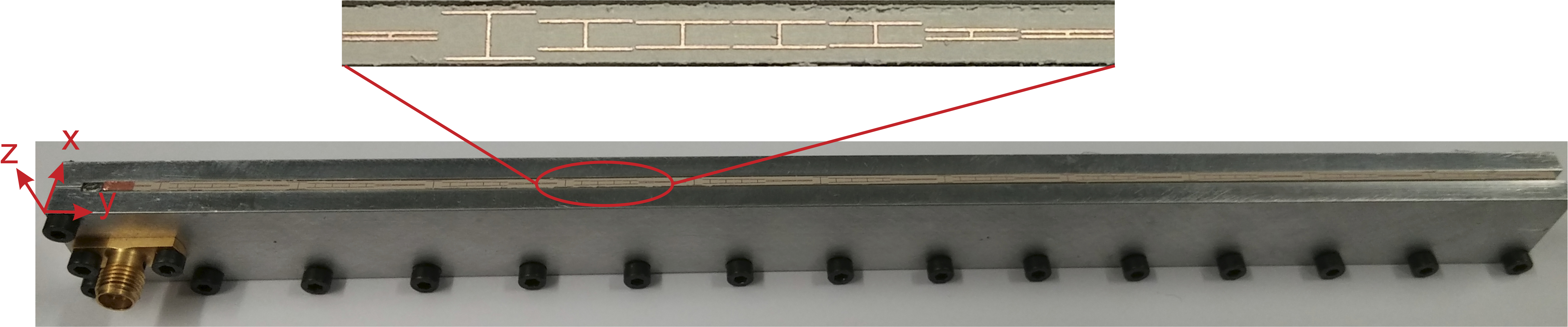}
\caption{Photograph of the fabricated LWA (Design 1).}
\label{fig:FotoPrototipo}
\end{figure}

Designs 1 and 3 have been manufactured and measured for experimental validation. As in the realistic full-wave simulations, three 50\,mil-thick laminates of RO3010 were used (four trace layers), bonded by Rogers 2\,mil-thick 2929 bondply. The metasurface fabrication was carried out by Candor Industries Inc. Several replicas of the same design were realized. The waveguide was fabricated on a 4\,mm Aluminium block at the University of Toronto and was manufactured in two pieces (split along the longitudinal axis) to facilitate the fabrication using computerized numerical control (CNC) technology. Then, the structure was assembled using metallic screws, as done in \cite{Ep16_Nat}. To feed the structure, a SMA connector with an exposed pin along the x-direction was used as a current source to excite the TE field. The whole structure with the connector was simulated in HFSS to choose the best distance to the back short in terms of matching at the design frequency (5.4\,mm). Fig. \ref{fig:FotoPrototipo} shows a photograph of the fabricated LWA corresponding to Design 1. 

The radiation patterns in the yz-plane were measured in the anechoic chamber of the University of Toronto in the frequency range from 18 to 22 GHz. Some absorbers were attached at the end of the metasurface to avoid radiation from the end. 

\begin{figure}[t!]
\centering
\includegraphics[width=0.9\columnwidth]{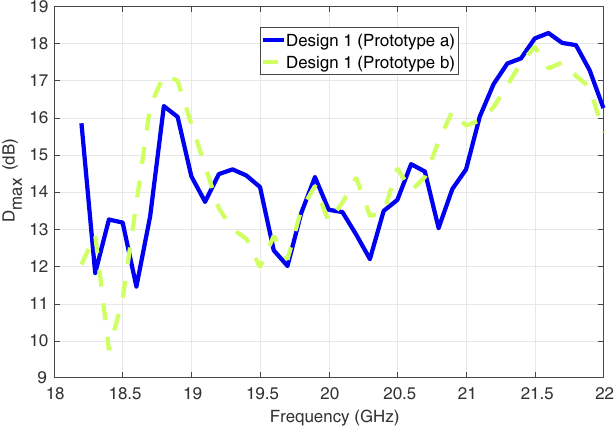}
\caption{Measured maximum 2D directivity of two replicas of Design 1 over frequency.}
\label{fig:MeasuredDmax}
\end{figure}

Fig. \ref{fig:MeasuredDmax} shows the maximum 2D directivity as a function of frequency for two replicas of the metasurface of Design 1, implying reasonable repeatability. It can be observed that the maximum directivity (which corresponds to the best mitigation of spurious Floquet modes) is found around 21.5 GHz, while the design frequency, as recalled, was 20 GHz. Therefore, there is a frequency shift in the expected behavior of the metasurface, which is attributed to fabrication tolerances and uncertainties in the actual value of the permittivity, as well as in the anisotropy of the dielectric.

\begin{figure}[t!]
\centering
\includegraphics[width=0.9\columnwidth]{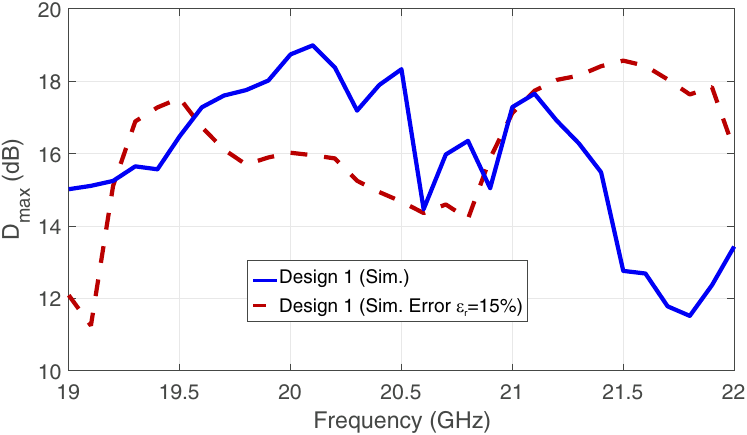}
\caption{Simulated 2D directivity of Design 1 over frequency with the permittivity value provided by the manufacturer and for a decrease in the permittivity by 15\%.}
\label{fig:SimulatedDmax}
\end{figure}

In order to verify that the frequency deviation can be attributed to deviations of the substrate permittivity from the nominal value provided by the manufacturer, Fig. \ref{fig:SimulatedDmax} shows the simulated 2D directivity vs. frequency when using the manufacturer provided value (solid blue line) and when considering a 15\% decrease in this value (dashed red line). It can be seen that a decrease in the permittivity leads to a shift of the frequency at which the maximum directivity is achieved. In particular, Fig. \ref{fig:SimulatedDmax} indicates that when considering 15\% permittivity deviation, the maximum directivity is achieved at 21.5\,GHz, coinciding with the peak performance point recorded in measurements (Fig. \ref{fig:MeasuredDmax}). Thus, it will be reasonable to conclude that the actual substrate permittivity is smaller by about 15\% from the value used for the design; from now on, we would use simulations with this modified value as the theoretical reference for the LWA performance.

\begin{figure}[t!]
\centering
\includegraphics[width=0.9\columnwidth]{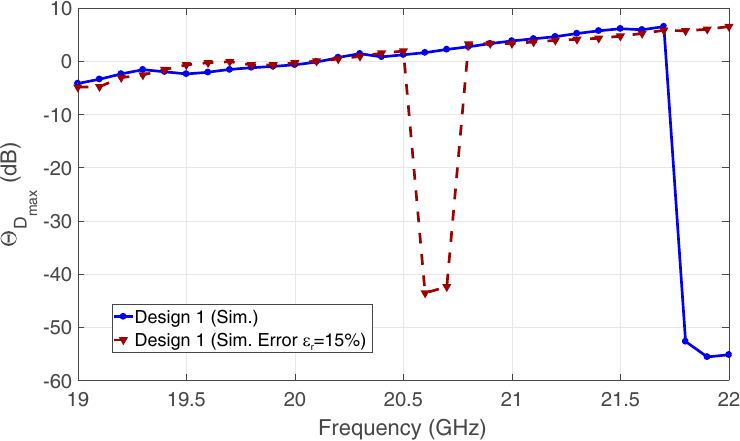}
\caption{Pointing angle over frequency obtained from full-wave electromagnetic simulations of Design 1 with the permittivity value provided by the manufacturer and for a decrease in the permittivity by 15\%.}
\label{fig:SimulatedThetaDmax}
\end{figure}

Assuming that the mode inside the waveguide is not affected much by the deviation in the permittivity of the metasurface substrate (so, $\theta_{in}$ is the one we designed), the pointing angle will be determined by the periodicity of the structure, given by (\ref{eq:period}). Subsequently, the relation between the frequency with the pointing angle is expected to be the same as in the design. Therefore, the frequency shift in the performance will lead to a pointing angle different from the designed one (broadside). Fig. \ref{fig:SimulatedThetaDmax} shows the pointing angle obtained from simulations of the designed case and the one with actual permittivity. It can be observed that the pointing angle practically does not change for the two cases, as assumed. 
The jumps at certain frequencies are due to the high level of the spurious Floquet harmonics (minimum values of directivities in Fig. \ref{fig:SimulatedDmax}) which makes the side lobe become the main one. With the assumed deviation in the permittivity, the boundary condition for the transformation of the guided field into the leaky-mode with mitigation of other Floquet modes is achieved at 21.5\,GHz, which, according to simulations, corresponds to around 5$^\circ$.

\begin{figure}[t!]
\centering
\includegraphics[width=7cm]{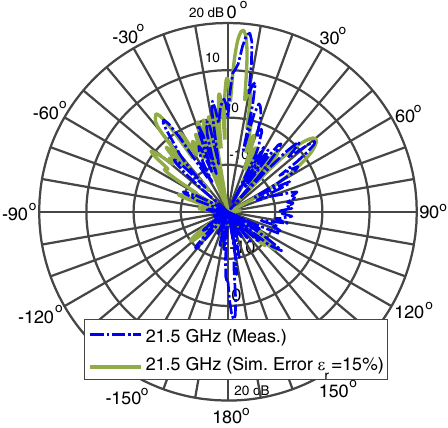}
\caption{Measured directivity pattern of Design 1 at 21.5\,GHz and the simulated directivity at 21.5\,GHz considering a permittivity deviation of $15\%$.}
\label{fig:MeasuredDirPattern}
\end{figure}

Fig. \ref{fig:MeasuredDirPattern} shows the directivity pattern in the zy-plane for the prototype `a' of Design 1 at 21.5\,GHz for both the measurement and the simulated directivity with the actual permittivity. It is shown that the mitigation of spurious Floquet modes is achieved with a side lobe level (SLL) of more than 14\,dB and 2D maximum directivity of 18.1\,dBi at $7^\circ$, with very good agreement with the simulation. Therefore, the initial discrepancies with the simulated design can be attributed to a deviation in the real permittivity with respect to the value provided by the manufacturer of the substrate. This can be expected since Rogers itself has proven that, besides the dielectric composition, there are other factors which influence the apparent permittivity of a substrate \cite{Co11} and the value they provide is valid under certain construction (i.e., microstrip) and frequency range. In any case, since spurious Floquet mode suppression has been achieved, the experiment verifies the concept. 
Moreover, in terms of directivity (18.1\,dBi, which corresponds to an aperture efficiency of $68.5\%$), the prototype results are in good agreement with the values shown in Table \ref{tab:PerformanceSummary}.

\begin{figure}[t!]
\centering
\includegraphics[width=\columnwidth]{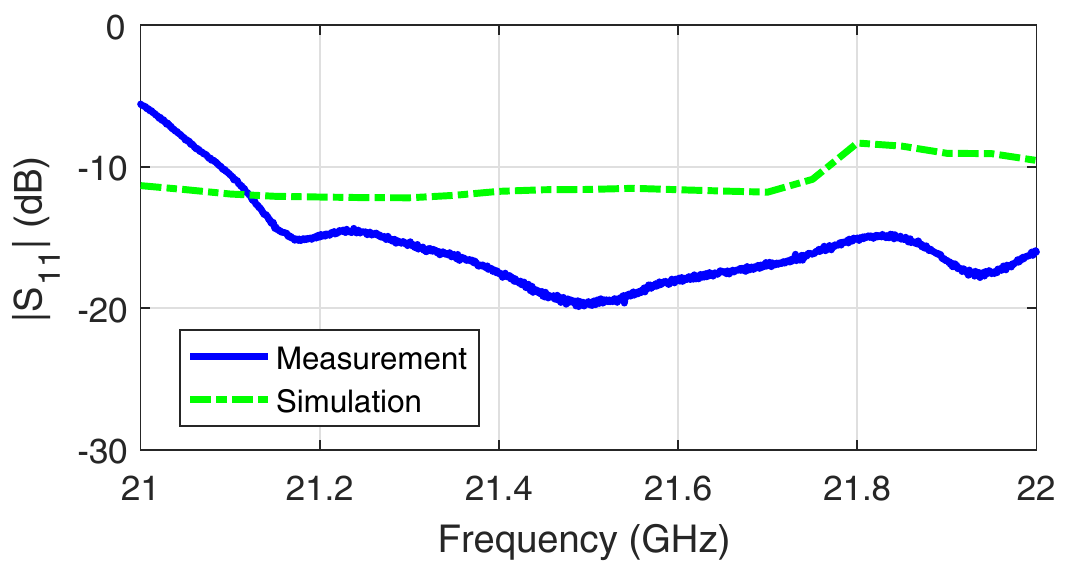}
\caption{Measured and simulated magnitude of the $S_{11}$ over frequency of Design 1. The simulation considers the connector and the permittivity deviation of $15\%$.}
\label{fig:MeasuredS11}
\end{figure}

Fig. \ref{fig:MeasuredS11} shows the measured $|S_{11}|$ for the prototype `a' of Design 1 together with the value obtained from the full-wave simulation when considering again the permittivity deviation of $15\%$ and the effect of the connector. Although the overall level of the $|S_{11}|$ is several dB lower in the measurement than in the simulation, the behavior is rather similar over most of the analysed band. The discrepancies can be attributed to fabrication tolerances, uncertainties in the actual connector dimensions and prototype assembly. Nevertheless, it is shown that the structure is well matched at the `shifted' design frequency (21.5\,GHz) both in the simulation and in the fabricated prototype. Additionally, the almost flat return loss behavior qualitatively agrees with the simulation of Fig. \ref{fig:S11_DesignA&C}, corroborating the smooth performance around the design frequency due to the excitation of only the radiating ($m=-1$) spatial harmonic.

\begin{figure}[t!]
\centering
\includegraphics[width=7cm]{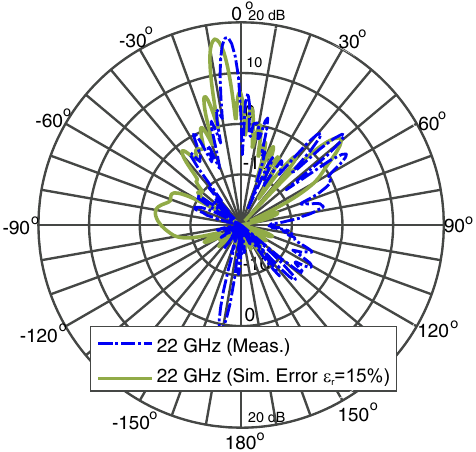}
\caption{Measured directivity pattern of Design 3 at 22\,GHz and the simulated directivity at 22\,GHz considering a permittivity deviation of $15\%$.}
\label{fig:MeasuredDirPatternDesignB}
\end{figure}

In the same way as done for Design 1, it can be shown that for the manufactured Design 3 the frequency at which the Floquet mode suppression is achieved has been shifted to 22 GHz. In order to check again if this shift is due to a deviation in the permittivity value of the substrate, Fig. \ref{fig:MeasuredDirPatternDesignB} shows the directivity pattern in the zy-plane for a prototype of Design 3 at 22\,GHz for both the measurement and the simulation when considering the same deviation in the permittivity of $15\%$. It can be observed that there is good agreement between the measurement and the simulation, thus corroborating the hypothesis. The small pointing deviation can be due to some minor miscalibration when performing the measurements. Additionally, it can be noticed that at 22\,GHz, the spurious Floquet modes are again mitigated (SLL lower than 10\,dBi, with maximum 2D directivity of 17.1\,dBi) and there is a single beam corresponding to the $m=-1$ harmonic. Due to frequency scanning, the pointing beam is not at -20$^\circ$ but the shift is in the expected direction, as for the prototype of Design 1. As expected, the pointing beam is backwards with respect to the one obtained for the prototype of Design 1, thus showing that other pointing directions can be achieved. Again, very good agreement in terms of directivity is obtained with the values shown in Table \ref{tab:PerformanceSummary}.

\section{Conclusion}
A novel concept of LWAs based on Omega-type bianisotropic Huygens' metasurfaces has been presented. The theoretical derivation has shown that there is an exact solution to convert a guided mode into a single radiating leaky mode by means of a periodic metasurface with arbitrary control of the leakage factor and the pointing beam direction. Although the solution for the metasurface is periodic, there is only one Floquet mode excited and radiating, even if the period is electrically large. Then, the traditional problem of the open-stopband effect at broadside due to coupling of Floquet harmonics does not appear.  

The theory has been verified with three different designs, two of them radiating at broadside with different leakage factors and a third one with a different pointing angle. Two types of simulations have been performed. First, idealized impedance sheets have been used in the simulator to implement the metasurface. In this way, the properties extracted from theory, independently of the microscopic design of the metasurface, have been checked. In a further stage, we have proposed physical structures for the three discussed designs, by following an implementation methodology in which several aspects of the physical realization have been discussed, such as the effect of the losses. Very good agreement has been verified between simulations and theoretical predictions. Furthermore, it has been illustrated that independent control of the phase and attenuation constants can be achieved. Moreover, broadside radiation has been shown without any degradation due to the open-stopband effect. Since the suppression of the spurious Floquet modes has also been achieved, the concept has been corroborated. 

Finally, the design cycle has been closed with the experimental verification of two prototypes. Deviations have been found in the frequency at which the Floquet mode suppression is achieved with respect to the design. It has been shown that this discrepancy is likely due to a deviation in the permittivity value with respect to the nominal value provided by the dielectric manufacturer. Nevertheless, it has been proven that a single beam with directivity close to the one obtained from simulations is achieved, which proves the control on the leakage factor and the spurious Floquet mode suppression even with structures with electrically long periods. 

It should be highlighted that our design relies completely on an analytical formulation of the fields, without any approximation. Therefore, it is expected that this work paves the way to implement LWAs with completely arbitrary radiation patterns.

\bibliographystyle{IEEEtran}

\begin{IEEEbiography}[{\includegraphics[width=1in,height=1.25in,clip,keepaspectratio]{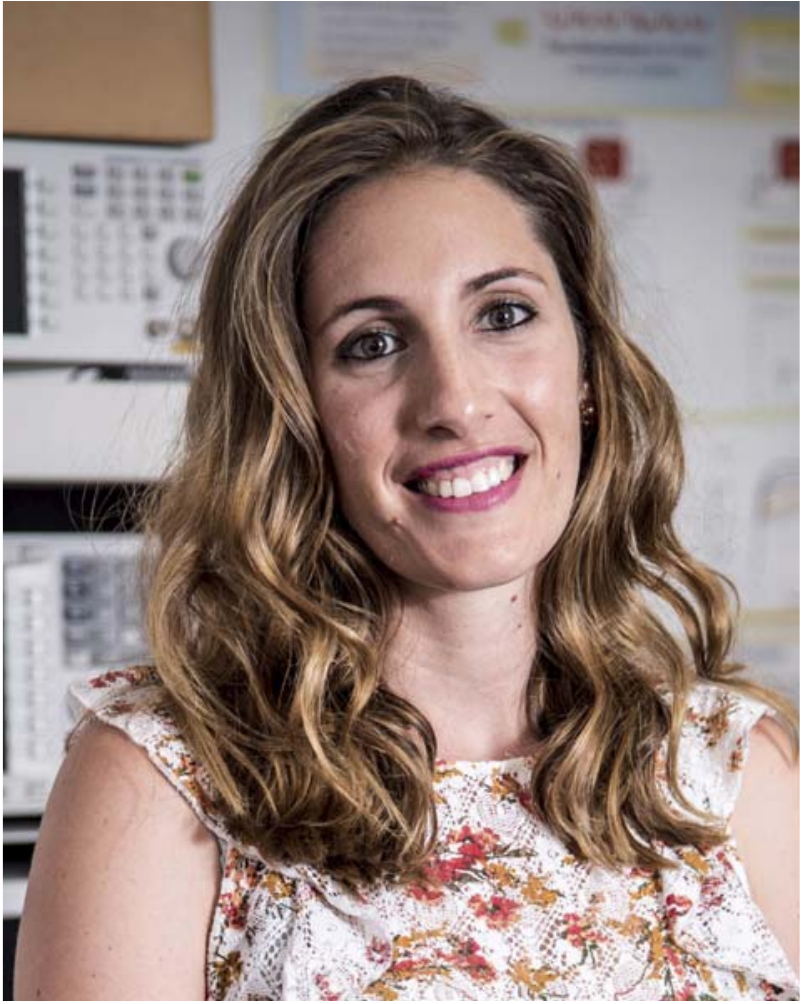}}]{Elena Abdo-S\'anchez}
(M'17) received the M.Sc. and Ph.D. degrees in Telecommunication Engineering in 2010 and 2015, respectively,  from the Universidad de M\'alaga, Spain. In 2009, she was a granted student with the Institute of Communications and Navigation at the German Aerospace Center (DLR) in Munich, Germany. In 2010, she joined the Department of Communication Engineering at the Universidad de M\'alaga as a Research Assistant. From April to July 2013, she was a visiting Ph.D. student at the Antennas and Applied Electromagnetics Laboratory of the University of Birmingham, UK. From May 2016 to May 2017, she was a Marie Sklodowska-Curie postdoctoral fellow at the Electromagnetics Group of the University of Toronto, Canada. She is currently a postdoctoral fellow at the University of M\'alaga. Her research interests focus on the electromagnetic analysis and design of planar antennas and the application of metasurfaces to the implementation of novel antennas.

Dr. Abdo-S\'anchez was recipient of both a Junta de Andaluc\'ia Scholarship (2012-2015) and a Marie Sklodowska-Curie fellowship (2016-2018).\end{IEEEbiography}

\begin{IEEEbiography}[{\includegraphics[width=1in,height=1.25in,clip,keepaspectratio]{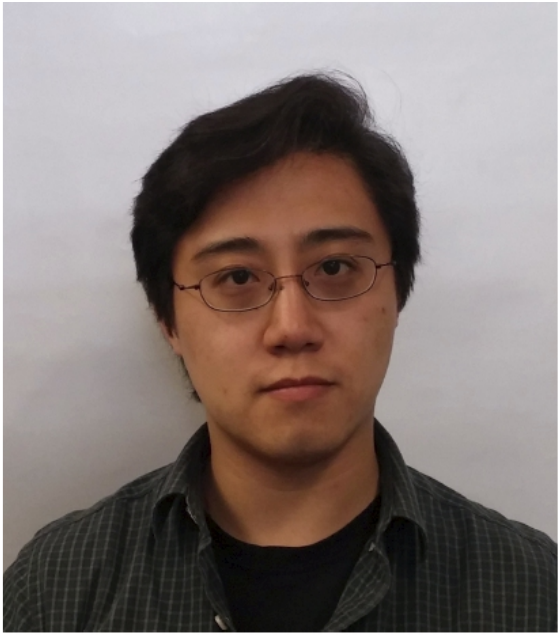}}]{Michael Chen}
(S'14) received the B.A.Sc. degree and M.A.Sc. degree in electrical engineering from the University of Toronto, Toronto, ON, Canada, in 2012 and 2015, respectively, and is currently working towards the Ph.D. degree at the University of Toronto. His research interests include antenna design, frequency-selective surfaces, metasurfaces, periodic structures, and radar systems.\end{IEEEbiography}

\begin{IEEEbiography}[{\includegraphics[width=1in,height=1.25in,clip,keepaspectratio]{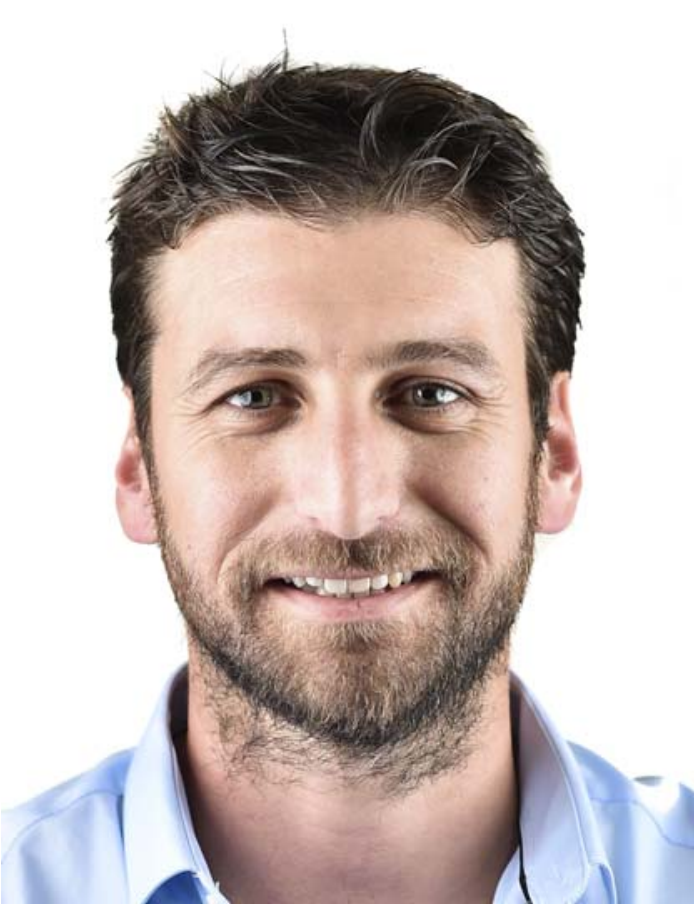}}]{Ariel Epstein}
(S'12-M'14) received the B.A. degree in computer science in 2000 from the Open University of Israel, Raanana, both the B.A. degree in physics and the B.Sc. degree in electrical engineering in 2003, and the Ph.D. degree in electrical engineering in 2013, all from the Technion - Israel Institute of Technology, Haifa, Israel. 
From 2013 to 2016, he was a Lyon Sachs Postdoctoral Fellow with the Department of Electrical and Computer Engineering, University of Toronto, Toronto, ON, Canada. He is currently an Assistant Professor with the Andrew and Erna Viterbi Faculty of Electrical Engineering, Technion - Israel Institute of Technology, Haifa, Israel, where he is leading the Modern Electromagnetic Theory and Applications (META) research group. His current research interests include utilization of electromagnetic theory, with emphasis on analytical techniques, for the development of novel metasurface-based antennas and microwave devices, and investigation of new physical effects.
Dr. Epstein was a recipient of the Young Scientist Best Paper Award in the URSI Commission B International Symposium on Electromagnetic Theory (EMTS2013), held in Hiroshima, Japan, in 2013, as well as the Best Poster Award at the 11th International Symposium on Functional $\pi$-electron Systems (F$\pi$-11), held in Arcachon, France, in June 2013. He is currently an Associate Editor for the IEEE Transactions on Antennas and Propagation.\end{IEEEbiography}

\begin{IEEEbiography}[{\includegraphics[width=1in,height=1.25in,clip,keepaspectratio]{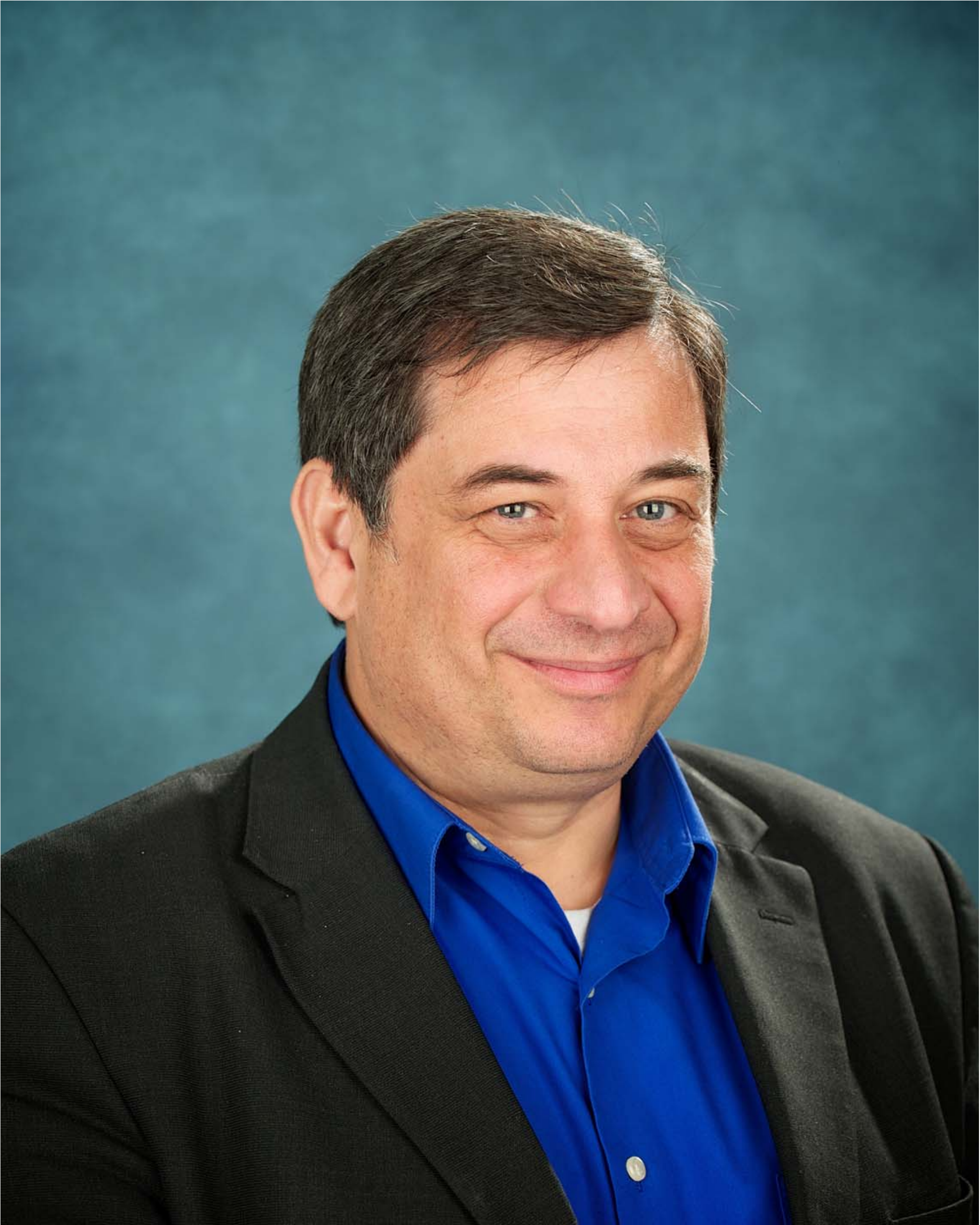}}]{George V. Eleftheriades}
(S'86–M'88–SM'02–F'06) earned the M.S.E.E. and Ph.D. degrees in electrical engineering from the University of Michigan, Ann Arbor, MI, USA, in 1989 and 1993, respectively. From 1994 to 1997, he was with the Swiss Federal Institute of Technology, Lausanne, Switzerland. Currently, he is a Professor in the Department of Electrical and Computer Engineering at the University of Toronto, ON, Canada, where he holds the Canada Research/Velma M. Rogers Graham Chair in Nano- and Micro-Structured Electromagnetic Materials. He is a recognized international authority and pioneer in the area of metamaterials. These are man-made materials which have electromagnetic properties not found in nature. He introduced a method for synthesizing metamaterials using loaded transmission lines. Together with his graduate students, he provided the first experimental evidence of imaging beyond the diffraction limit and pioneered several novel antennas and microwave components using these transmission-line based metamaterials. His research has impacted the field by demonstrating the unique electromagnetic properties of metamaterials; used in lenses, antennas, and other microwave and optical components to drive innovation in fields such as wireless and satellite communications, defence, medical imaging, microscopy, and automotive radar. Presently, he is leading a group of  graduate students and researchers in the areas of electromagnetic and optical metamaterials, and metasurfaces, antennas and components for broadband wireless communications, novel antenna beam-steering techniques, far-field super-resolution imaging, radars, plasmonic and nanoscale optical components, and fundamental electromagnetic theory. 

Prof. Eleftheriades served as an Associate Editor for the IEEE TRANSACTIONS ON ANTENNAS AND PROPAGATION (AP). He also served as a member of the IEEE AP-Society administrative committee (AdCom) from 2007 to 2012 and was an IEEE AP-S Distinguished Lecturer from 2004 to 2009. He served as the General Chair of the 2010 IEEE International Symposium on Antennas and Propagation held in Toronto, ON, Canada. Papers that he co-authored have received numerous awards such as the 2009 Best Paper Award from the IEEE MICROWAVE AND WIRELESS PROPAGATION LETTERS, twice the R. W. P. King Best Paper Award from the IEEE TRANSACTIONS ON ANTENNAS AND PROPAGATION (2008 and 2012), and the 2014 Piergiorgio Uslenghi Best Paper Award from the IEEE ANTENNAS AND WIRELESS PROPAGATION LETTERS. He received the Ontario Premier's Research Excellence Award and the University of Toronto's Gordon Slemon Award, both in 2001. In 2004 he received an E.W.R. Steacie Fellowship from the Natural Sciences and Engineering Research Council of Canada. In 2009, he was elected a Fellow of the Royal Society of Canada. He is the recipient of the 2008 IEEE Kiyo Tomiyasu Technical Field Award and the 2015 IEEE John Kraus Antenna Award. In 2018 he received the Research Leader Award from the Faculty of Applied Science and Engineering of the University of Toronto.
\end{IEEEbiography}

\end{document}